\documentclass[12pt]{article}

\RequirePackage[OT1]{fontenc}
\RequirePackage{amsthm,amsmath}
\RequirePackage[numbers]{natbib}
\RequirePackage[colorlinks,citecolor=blue,urlcolor=blue]{hyperref}

\usepackage[utf8]{inputenc}
\usepackage{float}
\usepackage{xcolor}
\usepackage{amsmath,amssymb}
\usepackage{mathabx}
\usepackage{graphicx}
\usepackage{lscape}
\usepackage{pifont}
\usepackage{fancyvrb}
\usepackage{hyperref}
\setcitestyle{authoryear,open={(},close={)}}

\numberwithin{equation}{section}

\newcommand{\R}{\mathbb{R}}
\newcommand{\E}{\mathbb{E}}
\newcommand{\pr}{\mathbb{P}}

\newcommand{\bD}{\mathbf{D}}
\newcommand{\bY}{\mathbf{Y}}
\newcommand{\by}{\mathbf{y}}

\newcommand{\bC}{\mathbf{C}}
\newcommand{\bZ}{\mathbf{Z}}
\newcommand{\bc}{\mathbf{c}}
\newcommand{\bz}{\mathbf{z}}
\newcommand{\bbeta}{\boldsymbol{\beta}}

\newcommand{\dd}{\mathrm{d}}

\newcommand{\PL}{\mathrm{PL}}
\newcommand{\CLRL}{\mathrm{CLRL}}
\newcommand{\WCLRL}{\mathrm{WCLRL}}
\newcommand{\BRLlogit}{\mathrm{BRL}_{\text{\scriptsize\rm logit}}}
\newcommand{\BRLcloglog}{\mathrm{BRL}_{\text{\scriptsize \rm cloglog}}}

\newcommand{\sub}{\mathrm{sub}}

\newcommand{\rev}[1]{{#1}}

\begin{document}


\title{Spatio-temporal point process intensity estimation using zero-deflated subsampling applied to a lightning strikes dataset in France}

\author{Jean-François Coeurjolly\footnote{Univ. Grenoble Alpes, CNRS, LJK, 38000 Grenoble, France; \texttt{jean-francois@coeurjolly@univ-grenoble-alpes.fr}}, Thibault Espinasse, \\Anne-Laure Fougères \footnote{Universite Claude Bernard Lyon 1, ICJ UMR5208, CNRS, Ecole Centrale de Lyon, INSA Lyon, Université Jean Monnet, 69622 Villeurbanne, France; \texttt{espinasse@math.univ-lyon1.fr, fougeres@math.univ-lyon1.fr}}, Mathieu Ribatet \footnote{Ecole Centrale de Nantes, Département Informatique et Mathématiques, Laboratoire de Mathématiques Jean Leray, 1, rue de la Noe 44321 Nantes Cedex, France; \texttt{mathieu.ribatet@ec-nantes.fr}}}

\date{} 




\maketitle

\begin{abstract}
Cloud-to-ground lightning strikes observed in a specific geographical domain over time can be naturally modeled by a spatio-temporal point process. Our focus lies in the parametric estimation of its intensity function, incorporating both spatial factors (such as altitude) and spatio-temporal covariates (such as field temperature, precipitation, etc.). The events are observed in France over a span of three years. 
Spatio-temporal covariates are observed with resolution $0.1^\circ \times 0.1^\circ$  ($\approx 100$km$^2$) and six-hour periods. 
This results in an extensive dataset, further characterized by a significant excess of zeroes (i.e., spatio-temporal cells with no observed events). We reexamine composite likelihood methods commonly employed for spatial point processes, especially in situations where covariates are piecewise constant. Additionally, we extend these methods to account for zero-deflated subsampling, a strategy involving dependent subsampling, with a focus on selecting more cells in regions where events are observed. A simulation study is conducted to illustrate these novel methodologies, followed by their application to the dataset of lightning strikes.
\end{abstract}





\section{Introduction}

Severe thunderstorms can be associated with personal injury or costly infrastructure damage. Among the major risks to be assessed are those associated with high lightning intensity, and they have been studied for about twenty years (see e.g. \cite{curran-etal2000, mona-etal2016, schulz-etal2005, schulz-etal2016, simon-etal2017,hernandez2019point,nampak2021characterizing}). Additionally, the interest in evaluating the intensity of cloud-to-ground lightning strikes might increase in the context of an energy transition: wind farms are multiplying, and the risk of lightning is one of the five hazards typically identified by manufacturers. From a physical standpoint, a thunderstorm corresponds to the dissipation of energy involving various thermodynamic processes, such as convection. To describe these complex processes, several covariates are known to be useful \rev{\citep[e.g.][]{bertram-mayr,chauduri-middey,taillardat-mestre2020}}: the CAPE (Convective Available Potential Energy), which acts as the 'fuel' in the development of the cloud; the $\theta_w'$ at 850 hPa (Wet-bulb potential temperature at 850 hPa), which is known to be a good synoptic predictor, giving important information on the position of the fronts; the temperature at two different altitudes (20m and 1500m), providing information on the vertical profile of the air mass (convection); humidity; and wind (zonal and meridional components). A few other purely spatial covariates like the altitude, the distance to the sea should also have an influence (see~Section 2 for more details). Lightning strike impacts, although continuously studied by scientists in the fields of physics, climatology or statistics remains a phenomenon with an important part of randomness. The set  of strikes collected over time and space constitute a spatio-temporal point pattern and the goal of this paper is to characterize its distribution in terms of climatological, topographical covariates.  

Spatio-temporal point processes are the stochastic models generating such data. Point processes in (a Polish) space $S$ \citep{daley2003introduction,moller2003statistical}
model points, events, objects in interaction. The data consist in a set $\by=\{y_1,\dots,y_n\}$ with $y_i\in S$. When $S=\R^+, \R^d$ ($d\ge 1$) or $\R^d\times\R^+$ we speak of temporal, spatial or spatio-temporal point processes. Most of statistical methodologies and their implementation have been devoted to temporal or spatial point processes \citep{moller2003statistical,illian2008statistical,baddeley2015spatial}. Spatio-temporal are more challenging and modern  due to the increase of data collection and have also received a lot of attention \citep{diggle2006spatio,cressie2015statistics,gonzalez2016spatio}. 

The aim of this paper is to explore first-order structure of the lightning strikes dataset. The intensity function, say $\rho(y)$ for $y\in \R^d\times \R^+$, measures the mean local number of events in the vicinity of a space-time point $y$. A point pattern is said to be homogeneous if the intensity is constant and inhomogeneous otherwise. 
Our objective is to model parametrically the function $\rho$ using spatial and spatio-temporal covariates, to estimate these parameters and to produce spatial prediction maps. This work can be viewed as a climatological risk assessment contribution. We characterize the distribution of lightning strikes using covariates. We do not indent to produce a short-term forecast of lightning strikes activity. 

Before detailing further our approach and contributions, let us first highlight a few facts about the dataset (detailed in Section~\ref{sec:desc}): we observe more than one million of impacts between 2013 and 2015, together with 6 spatial covariates and 7  spatio-temporal covariates. The spatio-temporal resolution is $v=0.1^\circ \times 0.1^\circ \times 6\text{h}$. As outlined in Section~\ref{sec:desc}, the point pattern is  highly inhomogeneous in time with daily and seasonal effects, and in space. In addition, the point pattern exhibit some (natural) clustering effect not explained by inhomogeneity. Together these last two facts are the reasons why approximately 99\% of cells of size $v$ do not contain any data point. 

We consider an exponential family model for the intensity function $\rho$, see Section~\ref{sec:background} and in particular Equation~\eqref{eq:rhoY}.
On the one hand, the clustering effect prevents us from modelling the data by a spatio-temporal Poisson point processes, the standard reference process which models events without any interaction (see e.g.~\citet{daley2003introduction}). On the other hand, maximum likelihood estimation is also known (and in particular for spatio-temporal point processes) to be very computationally expensive \citep{moller2003statistical}. Composite likelihood methods offer excellent alternatives. They were not all developed for spatio-temporal point processes but the extension is in principle straightforward. Among methods, we have the Poisson likelihood  \citep{rathbun1994asymptotic,baddeley2000practical,schoenberg2005consistent,waagepetersen2009two}, the  logistic regression likelihood when one focuses not on the number of impacts per cell but on the presence/absence of impacts, a standard approach known as spatial pixel-based logistic regression and commonly used in the standard GIS community \citep[see e.g.][and the many references therein on this topic]{baddeley2010spatial}, the conditional logistic regression likelihood estimation \citep{waagepetersen2008estimating,baddeley2014logistic}, the quasi-likelihood method \citep{guan2015quasi} or the variational method \citep{coeurjolly2014variational}. We review and compare these methods (except the last two ones which are less appropriate in the context of this application as explained in Section~\ref{sec:otherapproaches}) in Section~\ref{sec:estimation}. More precisely, we particularize the different methods and discuss implementation issues, in the context of discretized covariates (ie piecewise constant covariates on cells of volume $v$). As noticed and studied by~\cite{baddeley2010spatial}, many papers in environmental sciences, ecology and, say, research areas using data  based on Geographical Information Systems (GIS) analyse spatial point pattern data at a pixel level~\citep{elliot2000spatial,wakefield2004critique,waller2004applied}. Such a framework is often encountered in practice, see e.g. \citet{raeisi2021spatio}. Our analysis in Section~\ref{sec:estimation} yields interesting facts and links with generalized linear models. Some of these facts are known and can be found in \citet{baddeley2000practical,baddeley2010spatial,baddeley2014logistic}. In particular, we demonstrate that the Poisson likelihood method, acknowledged for experiencing slight bias in the presence of continuous covariates due to an approximation in quasi-Poisson regression implementation, is precisely equivalent to a Poisson regression with a specific offset term when covariates are discretized. 

The resolution of the covariates, combined with the fact that the observed phenomenon is highly inhomogeneous in time and space explain the overwhelming dominance of zero values of the indicators of presence/absence of an impact in a spatio-temporal cell. As detailed in Section~\ref{sec:desc}, whatever the season or six-hour periods, we observed that approximately 99\% of cells with volume $v$ do not contain any data point. When one applies any method described above to the dataset or to any independent subsample (to reduce computational cost inherent to this large dataset), this causes numerical problems and instabilities. A practical strategy that is common in the GIS literature~(\citet{atkinson1998generalised,gorsevski2006spatial}, see also \citet[Section 9.10.3]{baddeley2015spatial}) when pixel-based logistic regression is considered is to take a sample of pixels where zero values are observed. Then logistic regression is applied to the data consisting of the ‘1’ pixels (pixels containing at least one data point) and the sampled ‘0’ pixels. The main contribution of the present paper is to revisit standard composite likelihood-methods for point patterns (temporal, spatial or spatio-temporal) when such ideas are considered. It is worth specifying that we do not aim at modelling this excess of zeroes. Instead one aims at defining subsamples where the '0' (here) voxels can be deflated compared to the '1' voxels. Section~\ref{sec:subsampling} details the construction of such subsamples, called zero-deflated subsamples, and shows how these ideas can be applied to all composite likelihood methods presented in Section~\ref{sec:estimation}.

The rest of the paper is organised as follows. We detail the dataset and the problems it implies in Section~\ref{sec:desc}. Section~\ref{sec:background} provides a background on spatio-temporal point processes and specifies the parametric model considered in this paper. Section~\ref{sec:estimation} reviews composite likelihood methods used to estimate the intensity function in the particular case where covariates are discretized. Section~\ref{sec:subsampling} defines the concept of zero-deflated subsamples in particular when they are applied to methods defined in Section~\ref{sec:estimation}. We end Section~\ref{sec:subsampling} with a brief simulation study showing the efficiency of the proposed methodologies when they are applied to zero-deflated subsamples even if we have a dataset with a large number of '0'. Section~\ref{sec:backDataset} proposes an application to the lightning strikes dataset. We first present criteria to measure quality of spatio-temporal predictions, one of them being very close to the Wasserstein distance largely used in the statistics community to compare two empirical measures. Second, we compare methods described in Section~\ref{sec:subsampling} in terms of computational cost and prediction quality. Our finding is that the subsampled version of the Poisson likelihood turns out to be the best one. This method is then used to illustrate temporal prediction curves and spatial prediction maps. Several perspectives can be addressed based on the present work. Some of them are described in Section~\ref{sec:conclusion}. Finally, Appendix~\ref{app:figures} gathers additional figures regarding the exploratory data analysis and simulation study.

\section{Presentation of data and their complexity}\label{sec:desc}

\begin{figure}[htbp] 
	\centering
\includegraphics[width=.6\textwidth]{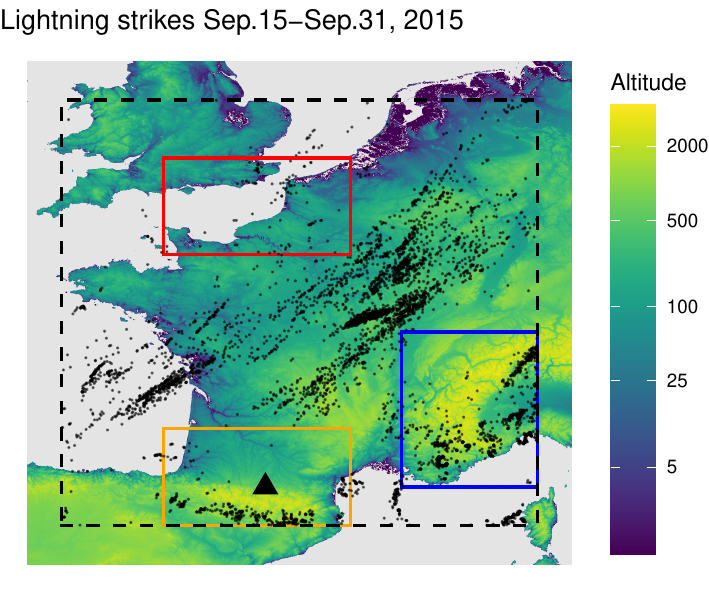}

\bigskip

\begin{tabular}{cccc}
\includegraphics[width=0.28\textwidth]{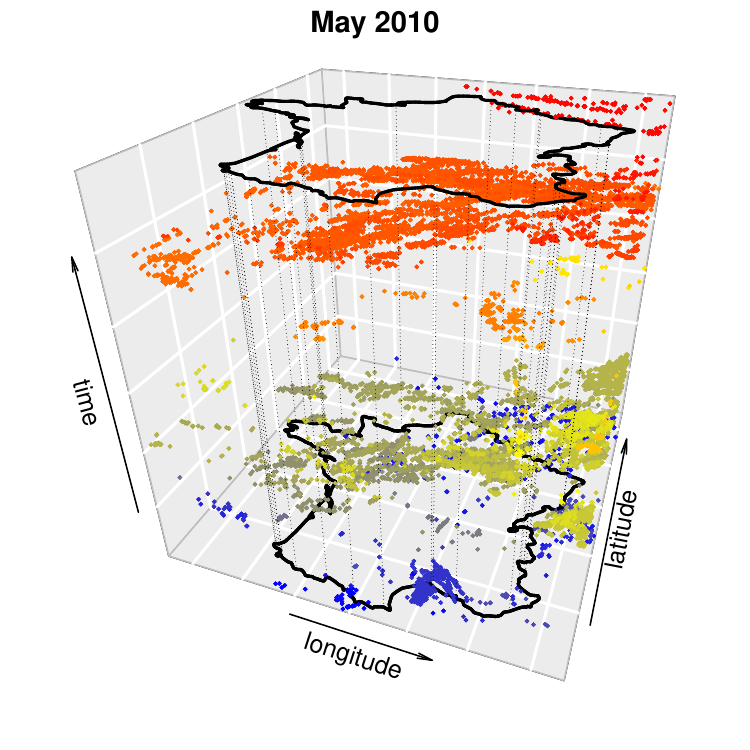}&
\includegraphics[width=0.28\textwidth]{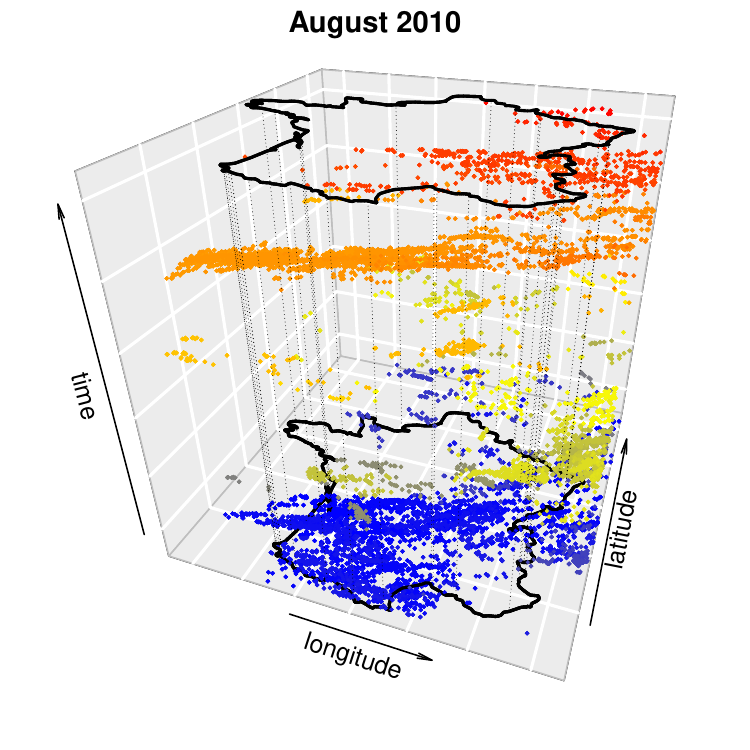}&
\includegraphics[width=0.28\textwidth]{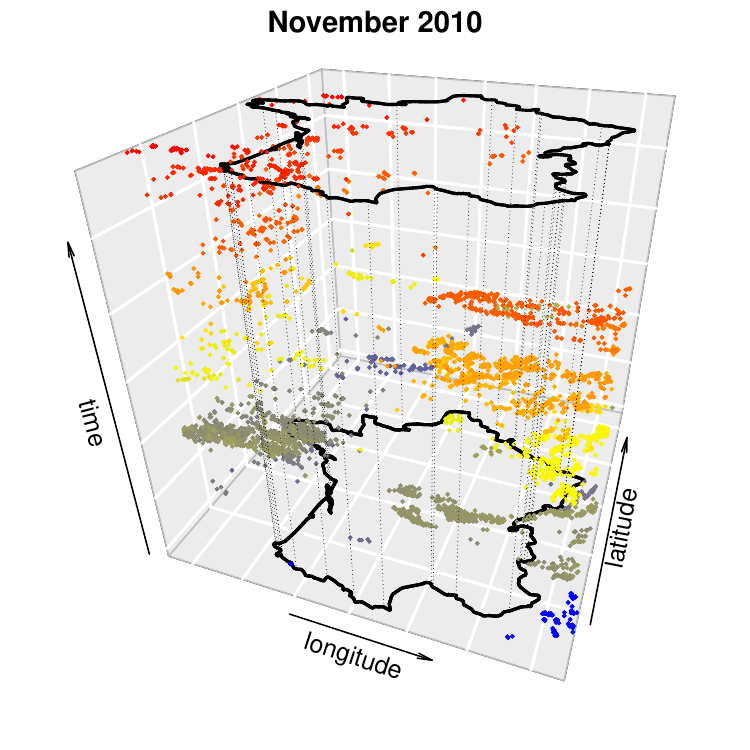}
\end{tabular}
\caption{(Top) Map of France with altitudes and locations of impacts during the second half of September 2015. The dashed black box (resp. blue, red, orange) corresponds to the  domain where impacts are observed (resp. to the subdomain called `Alps', `Channel' and `Pyrénées' in Appendix~\ref{app:eda}). 
The black triangle corresponds to the location denoted by $x_0$ in Figure~\ref{fig:covar}. 
(Bottom) Spatio-temporal representation of  lightning strikes impacts for different months in 2010. Only 20\% of (randomly chosen) impacts are represented for August 2010.}
\label{fig:introductionData}
\end{figure}

In this section, our objective is to shed the light on the challenges the data of interest impose in terms of statistical methodology. 
Data are provided by Météorage and Météo-France. We observe impacts of cloud-to-ground lightning flashes  over France from 2010 to 2015. Longitudes and latitudes are collected to a 100m accuracy and time events upto the nanosecond. Overall, we observe 1846533 events.  Figure~\ref{fig:introductionData} is a brief illustration of this spatio-temporal dataset, from which we can readily see that this phenomenon turns out to be highly inhomogenous in time and space. 

Even if a more rigorous background on point processes and intensity functions is provided in Section~\ref{sec:background}, Figures~\ref{fig:countingProcess} and~\ref{fig:impacts-mois} are quite easy to interpret. Figure~\ref{fig:countingProcess} focuses on the temporal nature of the dataset (by aggregating data over space). A yearly seasonal effect is clearly seen with a much higher intensity during summer (June to August) than in spring (April to May) and autumn (September to November). The winter period (December to March) seems to be the most quiet period. As shown in~Figure~\ref{fig:countingProcess} (bottom), the temporal intensity is quite variable for different regions (Alps, Channel, Pyrénées for instance). Finally, it is worth pointing out that within a day the temporal intensity varies a lot. We notice a slow activity during the night and a higher activity in the afternoon/evening, a pattern which occurs quite similarly for the different considered regions and seasons. Figure~\ref{fig:impacts-mois} is more focused on the spatial nature of the dataset. We aggregate data by month and estimate non parametrically the spatial intensity (using a standard kernel intensity estimation method, e.g. \cite{diggle1985kernel,moller2003statistical}). We retrieve the previous comment that the activity is higher during warmer months. This figure illustrates more strenuously the spatial differences on the French territory.

\begin{figure}[htbp]
 \centering
\includegraphics[width=\textwidth]{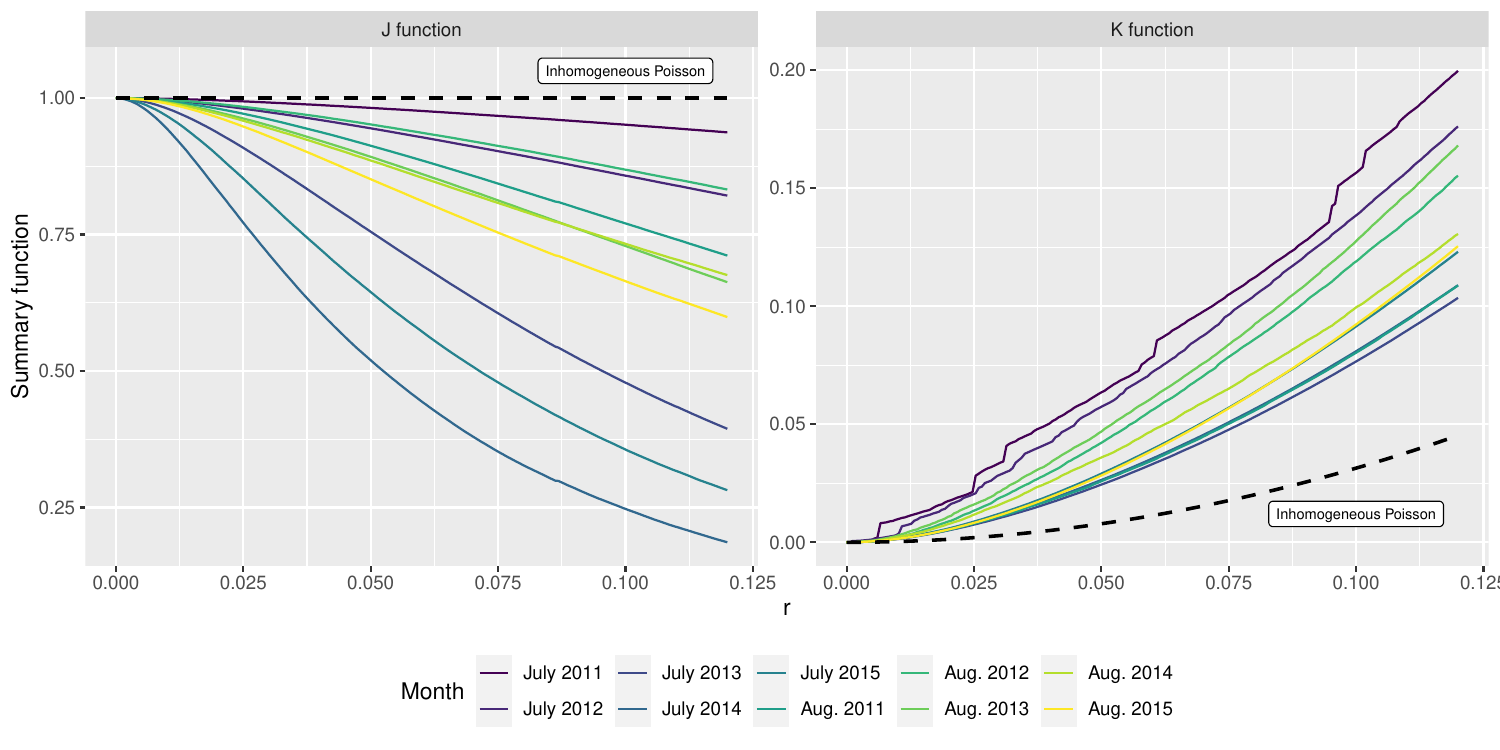}
\caption{Non parametric estimation of the spatial inhomogeneous $J$ and Ripley's $K$ functions. Data are aggregated by months for July and August from 2011 to 2015. The dashed line and  curve correspond to the expected $J$ and $K$ functions under the Poisson assumption. An estimate of the $J$ function (resp. $K$ function) below 1 (resp. above $\pi r^2$) which corresponds to the situation under the Poisson case indicates clustering in data, that is strong positive spatial correlation between events (ie lightning strikes).}
\label{fig:KJ}
\end{figure}

Besides the obvious inhomogeneous nature of this dataset, Figure~\ref{fig:KJ} explores the dependence between lightning strikes. This constitutes only a part of a larger exploratory analysis.  Here, we have  aggregated data by month and have estimated standard spatial summary statistics such as the inhomogeneous $K$ and $J$ functions (see \cite{moller2003statistical,illian2008statistical,van2011aj}). To estimate these functional summary statistics, we use kernel intensity estimates as plug-in estimates. We do not provide global envelopes tests (\cite{myllymaki2017global}) to not overload the graphs. The general conclusion is however crystal-clear: the point pattern cannot be simply modelled by a spatio-temporal inhomogeneous Poisson point process (otherwise monthly aggregated spatial point processes would be spatial Poisson point processes as well, which is not the conclusion from Figure~\ref{fig:KJ}). It is not the objective of the present paper to model second-order characteristics of the lightning strikes dataset. However, the previous comment means that if we  estimate parametrically the intensity of the point pattern, we have to consider an estimation method which is robust to the model.

In addition to this spatio-temporal point pattern, we also have at our disposal spatial covariates and
spatio-temporal covariates used to explain the distribution of lightning strikes. 
Spatial covariates are observed on a regular grid of $0.1^\circ \times 0.1^\circ$ ($\approx 100$km$^2$). 
They are constituted by the altitude map, the distance to the sea, as well as four additional variables computed from the AURELHY method \citep{benichou1994}. Their interest has been shown recently, see e.g.~\cite{taillardat-mestre2020}, for statistical post-processing of ensemble forecasts.
Spatio-temporal covariates consist of model outputs (Météo-France ARPEGE\footnote{See e.g. \href{https://www.umr-cnrm.fr/spip.php?article121&lang=en}{https://www.umr-cnrm.fr/spip.php?article121\&lang=en}} Model output) of climate covariates that are available only from 2013 to 2015. We have access to the following covariates observed on a regular grid of $0.1^\circ \times 0.1^\circ \times 6\text{h}$: CAPE (Convective Available Potential Energy); Humidity (at altitude 1500m); Temperatures (at altitudes 20m and 1500m);  $\theta_w'$ at 850 hPa (Wet-bulb potential temperature at 850 hPa); zonal and meridional  components of wind. The relevance of these latter covariates has been detailed in the Introduction.
The altitude map is illustrated in Figure~\ref{fig:introductionData}. Similarly to the point pattern, these covariates obviously vary a lot, as shown in Figure~\ref{fig:covar}. Figure~\ref{fig:pairplots} explores the interest of some of these covariates to explain the presence/absence of a lightning strike at a specific location. \rev{Maps presented in this paper do not take into consideration the earth curvature (a problem frequently pointed out in the literature \citet[e.g.][]{jeong_jun}), in particular because spatial and spatio-temporal covariates provided by Météo-France already consist on grid-projected data. 
}

\begin{figure}[htbp]
 \centering
\includegraphics[scale=.5]{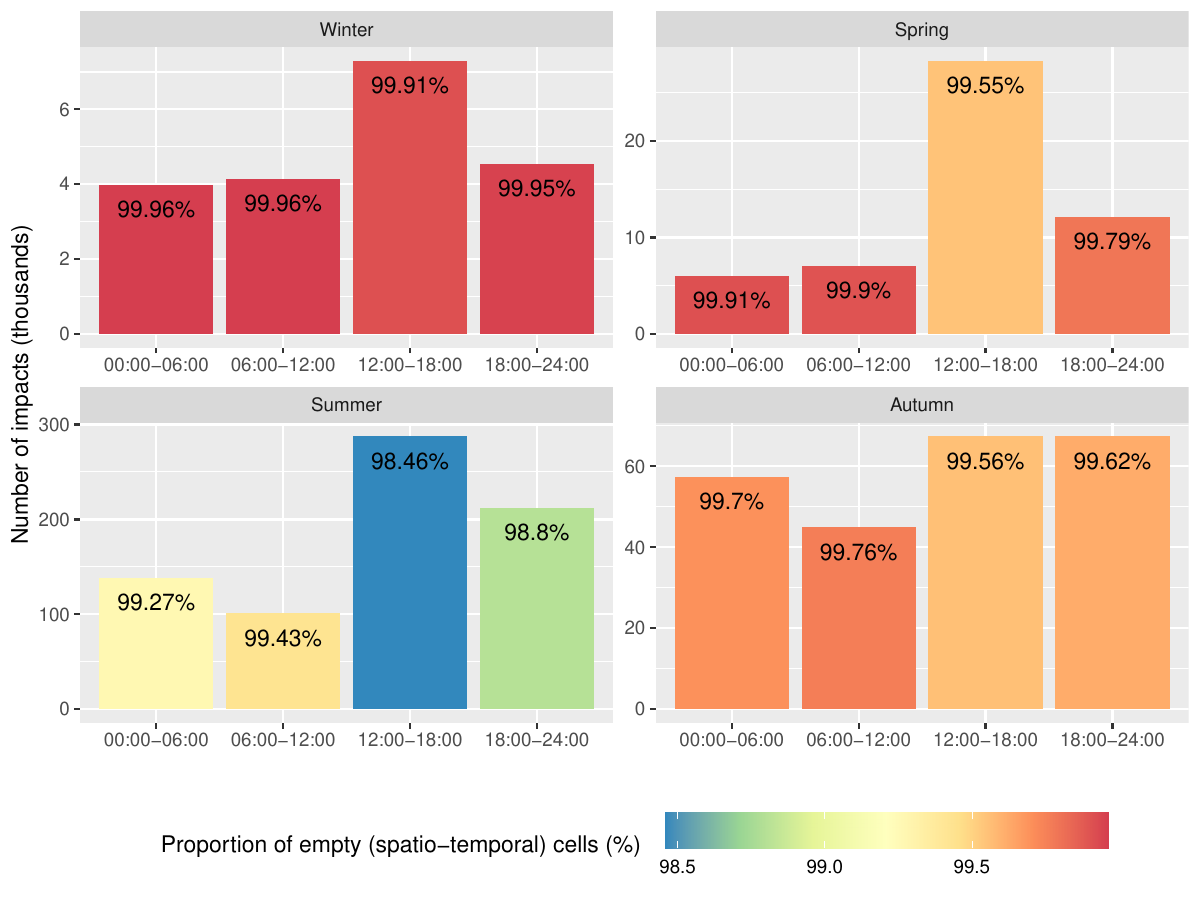}
\caption{Number of impacts (per thousands) by season and 6-hour period and proportion of the spatio-temporal voxel-grid cells containing no lightning strike impact. }
  \label{fig:zeroes}
\end{figure}

As noticed and studied by~\cite{baddeley2010spatial}, many papers in environmental sciences, ecology and, say, research areas using data  based on Geographical Information Systems (GIS) analyse spatial point pattern data at a pixel level. Such a framework is often encountered in practice, see e.g. \citet{raeisi2021spatio} and in particular in the current paper, as spatio-temporal covariates are collected within a 0.1 degree squared and a 6-hour period. Given the observed phenomenon (remind that data are collected up to the nanosecond), this spatio-temporal grid appears quite coarse. The spatio-temporal observation domain $W\times \mathcal T$ (where $W$ is the  spatial domain and $\mathcal T =[2013,2015]$) can therefore be viewed as a tessellation of spatio-temporal cells $\Delta_j$ ($j=1,\dots,J$) over which covariates are constant. Given the resolution of spatio-temporal covariates, we have $J \approx 7 \times 10^7$.
The following comment is crucial and constitutes the core of the present paper: despite the apparent high number of events (approximately $10^6$ lightning strikes impacts for data restricted to the period 2013-2015 where all covariates and spatio-temporal covariates are observed), most of the cells $\Delta_j$ do not contain any data. This is illustrated by Figure~\ref{fig:zeroes}: even if one splits data by season and 6-hour periods, more than 98\% of cells $\Delta_j$ are empty.

To summarize this section, we observe  1050342 spatio-temporal events of cloud-to-ground lightning strikes in France  over the period 2013-2015, 6 discretized spatial covariates and 7 discretized spatio-temporal covariates. The point pattern is highly inhomogeneous both in time and space and the events also seem to exhibit some strong  clustering (hence a departure to the independence assumption). Finally, at the voxel-grid level (i.e. at the resolution level), the dataset exhibits a huge excess of zeroes, in the sense that most of space-time observation domain is empty. All these different observations are studied in the next three sections before going back to the dataset in Section~\ref{sec:backDataset}.


\section{Notation and intensity modelling} \label{sec:background}

A (planar) spatio-temporal point process $\bY$ defined in $\R^2\times \R^+$ can be viewed as a locally finite random measure on $\R^2\times \R^+$ (see e.g. \citet{daley2003introduction,moller2003statistical}), which means that for any bounded $W\subset \R^2$ and $\mathcal T \subset \R^+$, the number of (space-time) points is bounded. We also assume that $\bY$ is simple, in the sense that two events do not occur at the same space-time point. In the following, we use the terminology point for an element of $\bY$, thus for a space-time point, and  the notation $y=(x,t)$ where $x\in W$ and $t\in \mathcal T$ for an element of $\bY$.

Estimating first-order summary statistics is usually the first step to analyse a point process. The first-order intensity function $\rho$ of $\bY$ (assumed to exist) is a function of $y=(x,t)$ and is defined by the following integral characterization (also called Campbell theorem), see \cite{moller2003statistical}: for any non-negative function $f$,
\[
  \mathbb E  \sum_{y \in \bY}  f(y) = \int_{\mathbb R^2 \times \mathbb R^+} f(y) \rho(y) \dd y
\]
where we abuse notation by denoting $\dd y=\dd x \dd t$. In other words, using the notation $N(A)$ for the counting variable of any bounded borel set $A\subset \mathbb R^2\times \mathbb R^+$
\[
  \rho(y) = \rho(x,t) = \lim_{\dd x\to 0, \dd t\to 0} \frac{\mathbb E\{N(\dd x \times \dd t)\}}{\dd x \dd t}
\]
describes the local number of points in the vicinity of $y=(x,t)$. Higher-order intensity functions can be described in a similar way (using higher-order version of Campbell theorem). In this paper, we are interested in modelling $\rho$ as a parametric form of spatial and spatio-temporal covariates (which themselves are realizations of spatial or spatio-temporal random fields) denoted here by $\bZ(x) \in \R^{n_Z}$ and $\bC(x,t) \in \R^{n_C}$. Hence, $n_Z$ and $n_C$ stand for the number of spatial and spatio-temporal covariates.  Given $\bC(x,t)=\bc$ and $\bZ(x)=\bz$, we model $\rho$ as a function of $x,t,\bc,\bz$, i.e. $\rho(y)=\rho((x,t),\bc,\bz)$. 

Given the explanatory study proposed in Section~\ref{sec:desc},  we propose to `aggregate' data by season and 6-hour period of the day. Following this, we let $\mathcal S=\{$summer, fall, winter, spring$\}$ and $\mathcal M=\{1,\dots,4\}$ denote the set of seasons (as described in the previous section) and the set of six-hour periods (0:00-06:00, 06:00-12:00, 12:00-18:00 and 18:00-0:00) of the day. And we view $\bY$ as a union of marked point processes where the mark is an element of $\mathcal S\times \mathcal M$ (which actually depends on time), that is $\bY = \bigcup_{(s,m)\in \mathcal S\times \mathcal M } \bY_{s,m}$ where $\bY_{s,m}$ can be described as 
$\bY_{s,m}=\{(x,t) \in \bY: x\in W \text{ and } t \in \mathcal T_{s,m}\}$ where $ \mathcal  T_{s,m}$ is the union of time intervals restricted to the season $s$ and the six-hour period $m$. Note that we do not assume that $\bY_{s,m}$ is independent of $\bY_{s^\prime,m^\prime}$. However, we assume that the intensity function $\rho$ of $\bY$ has the form
\begin{equation}\label{eq:rhoY}
  \rho(y) = \sum_{s\in \mathcal S, m\in \mathcal M} \rho_{s,m}(y)  \quad \text{ with } \quad
  \rho_{s,m}(y)=\exp\left( \bbeta_{s,m}^\top \left\{\bZ(x),\bC(x,t)\right\}\right)
\end{equation}
for any $y=(x,t)\in W\times \mathcal T_{s,m}$ with $\bbeta_{s,m}\in \mathbb{R}^{n_Z+n_C}$.
We do not make any further assumption on the spatio-temporal point process. In particular, we have to keep in mind that $\bY_{s,m}$ is probably not a Poisson point process, as revealed by Figure~\ref{fig:KJ}.

The estimation of $\rho$ from~\eqref{eq:rhoY} is tackled by estimating separately $\rho_{s,m}$ for any $s,m$. In the  following two Sections~\ref{sec:estimation}-\ref{sec:subsampling}, we focus on the specific problem of a single intensity. To make it general, we slightly simplify the notation by discarding the subscripts $s,m$ to ease the reading and omitting (without loss of generality) the spatial covariates $\bZ$. These sections are therefore written for general spatio-temporal point processes $\bY$ with intensity modelled from an exponential family model with discretized spatio-temporal covariates. 


\section{Composite likelihoods  methods with discretized covariates} \label{sec:estimation}

Let $\bY$ be a spatio-temporal point process on $W\times \mathcal T$. We consider the exponential model for its intensity given by $\rho(y)=\rho(x,t)=\exp\left( \bbeta^\top \bC(x,t)\right)$ for any $y=(x,t)\in W\times \mathcal T$, where $\bC(x,t)\in \R^{n_C}$. In particular,
we assume that the covariates are deterministic (to ease the presentation) and are constant on $\Delta_j$, that is $\bC(x,t)=\bC_j \in \R^{n_C}$ whenever $(x,t)\in \Delta_j$, where the cells $\Delta_j$, with volume $\delta_j$, form a tessellation of $W\times \mathcal T$, that is $W\times\mathcal T = \bigcup_{j=1}^J \Delta_j$. Note that the statements of Sections~\ref{sec:estimation}-\ref{sec:subsampling} are obviously valid for purely temporal or spatial point processes, or more generally for any point process in $\mathbb R^d$. 

The methods we consider fall into the terminology of composite likelihoods methods and are  quite standard in the literature, see e.g. \citet{coeurjolly2019understanding} for a review. However we particularize them and discuss implementation issues, in the context of discretized covariates, which yields interesting facts. Some of these facts can be found in \citet{baddeley2000practical,baddeley2010spatial,baddeley2014logistic}. A summary and comparison of  methods described below is proposed in Table~\ref{tab:summary}.

\subsection{Poisson likelihood (method $\PL$)} \label{sec:PL}

To estimate the vector $\bbeta$, it is now well-known (see e.g.~\cite{waagepetersen2009two} or~\cite{coeurjolly2019understanding}) that the maximum of the Poisson likelihood exhibits interesting properties such as consistency, asymptotic normality even if the underlying point process is not a Poisson point process. The main argument comes from Campbell theorem which in particular states that the score of the Poisson likelihood is unbiased for general point processes. For general intensity models, the Poisson likelihood writes
\begin{equation}\label{eq:PL}
  \PL(\bbeta) = \sum_{y\in \bY} \log \rho(y) - \int_{W\times \mathcal T} \rho(y)\dd y
\end{equation}
which, for discretized covariates and exponential family models~\eqref{eq:rhoY}
, reduces to
\begin{align}
\PL(\bbeta) &= \sum_{j=1}^J\left\{
N_j \bbeta^\top \bC_j - \delta_j \exp(\beta^\top \bC_j)
\right\} \nonumber \\
&=  \sum_{j=1}^J  
\left\{
N_j \log (\delta_j \exp(\bbeta^\top \bC_j)) - \delta_j \exp(\bbeta^\top \bC_j)
\right\} -  \sum_{j=1}^J N_j\log \delta_j \; .\label{eq:PL2}
\end{align}
The first sum of~\eqref{eq:PL2} precisely corresponds to a Poisson regression likelihood with  the canonical link and with  offset term $(\log \delta_j)_j$. Therefore, when covariates are piecewise constant, the Poisson (or composite) likelihood estimator is strictly equivalent to a Poisson regression (with an offset term). 

It is worth mentioning that when covariates are not piecewise constant, a  numerical problem appears since one has to discretize the integral term of~\eqref{eq:PL}. \citet{baddeley2000practical} use Berman-Turner's approximation \citep{berman1992approximating} and show that if one uses as quadrature points the union of a set of grid points and data points, Equation~\eqref{eq:PL} can be implemented via a quasi-Poisson regression model. However, using data points in the integral discretization step leads to an approximation which can be problematic especially when the number of data points is high, as noticed by~\citet{baddeley2014logistic}. 

\subsection{Conditional logistic regression and its weighted version (methods $\CLRL$ and $\WCLRL$)} \label{sec:CLRL}

When covariates are continuous, the implementation of the Poisson likelihood induces some bias. To overcome that numerical problem, \cite{waagepetersen2008estimating,baddeley2014logistic} propose an alternative estimating function which can be implemented using a logistic procedure. 
Following \citet{waagepetersen2008estimating}, on can consider a dummy point process given by a stratified binomial point process with $J$ points, one in each cell $\Delta_j$, that is the point process, say $\bD$, with 
$J$ points and with intensity $\eta(x)=\sum_{j=1}^J \delta_j^{-1}1(x\in \Delta_j)$. We then leave the reader to check that the conditional logistic regression likelihood (for spatial point processes) is equal to
\begin{equation}\label{eq:clrl}
\CLRL(\bbeta)=
\sum_{y\in \bY} \log\left( \frac{\rho(y)}{\eta(y)+\rho(y)}\right) + \sum_{y\in \bD} \log\left(\frac{\eta(y)}{\eta(y)+\rho(y)} \right).
\end{equation}
Equation~\eqref{eq:clrl} is indeed a logistic regression likelihood with a sample of $N+J$ Bernoulli random variables with probability $\pi(y)=\rho(y)/(\eta(y)+\rho(y))$ if $y \in \bY$ and $1-\pi(y)$ if $y\in \bD$. Let us see how~\eqref{eq:clrl} rewrites for piecewise constant covariates. Let $\bY=\{y_1,\dots,y_N\}$,  $\bD=\{d_1,\dots,d_J\}$ and $\rho_j=\exp(\beta^\top \bC_j)$, for $j=1,\dots,J$. For any sequence $a_1,\dots,a_{N}$, we define the sequence $\check a_1,\dots,\check a_{N+J}$ by
\[
  \check a_k = \left\{
  \begin{array}{ll}
   \sum_{l=1}^N a_l 1(y_k\in \Delta_l)& \text{ if } k=1,\dots,N \\
   a_{k-N} & \text{ if } k= N+1,\dots,J
  \end{array}
  \right.
\]
Using this notation, \eqref{eq:clrl} can be rewritten as
\begin{equation}\label{eq:clrl2}
 \CLRL(\bbeta)= \sum_{k=1}^{N+J} 
\left\{
\check i_k \log\left( \frac{\check \delta_k\check \rho_k}{1+\check \delta_k\check \rho_k}\right) 
+ (1-\check i_k)\log\left(\frac{1}{1+\check \delta_k\check \rho_k} \right)
\right\}
\end{equation}
where $\check i_k=1$ if $k=1,\dots,N$ and 0 otherwise. Formally, the latter equation corresponds to a logistic regression with Bernoulli random variables $\check I_k$ with mean $\mathrm P(\check I_k=1)={\check \delta_k\check \rho_k}/({1+\check \delta_k\check \rho_k})$. \citet{baddeley2014logistic} show that, for general spatial covariates and for any spatial point process, this procedure leads to an accurate estimation which is robust to the Poisson assumption. There is no integral discretization. It is also shown that if $J$ is large, the conditional logistic regression likelihood tends almost surely to the Poisson likelihood. Finally, since covariates are constant on $\Delta_j$, it is worth noting that~\eqref{eq:clrl2}, which is a logistic regression with response vector of length $N+J$, can be rewritten as
\begin{equation} \label{eq:clrl3}
\WCLRL(\bbeta)= \sum_{j=1}^{\# J_1+J} \tilde w _j \left( 
\tilde I_j \log(\tilde p_j) + (1-\tilde I_j)\log(1-\tilde p_j)
\right)
\end{equation}
where $\# J_1$ is the number of elements of $J_1 = \{j: N_j>0\}$ and where 
$$
(\tilde w_j,\tilde I_j,\tilde p_j)=\left\{
\begin{array}{ll}
 (N_{j_1(j)}, 1, p_{j_1(j)}) & \text{ if } j\le \#J_1   \\
 (1, 0, p_{j-\#J_1+1}) & \text{ if } J_1<j \le \#J_1+J
\end{array}
\right.
$$
where we use the notation $j_1(j)$ for the $j$th index of $\{1,\dots,J\}$ for which $N_j>0$. Equation~\eqref{eq:clrl3} can be simply implemented as a weighted logistic regression likelihood with canonical link. This implementation is faster since the response vector of the regression~\eqref{eq:clrl3} is now $\#J_1+J$ which can be much smaller than $N+J$ when a large number of cells $\Delta_j$ do not contain any point.

\subsection{Bernoulli regression likelihood with logit link (method $\BRLlogit$)} \label{sec:BRLlogit}

We continue with a method often considered in the literature (see e.g. \cite{baddeley2010spatial}), known as the pixel logistic regression. Let us start with Equation~\eqref{eq:clrl2}, which can  be written using~\eqref{eq:clrl} as
\begin{equation} 
 \sum_{j=1}^J \left\{
N_j \log\left( p_j\right) 
+ \log\left(1-p_j \right) 
\right\} \quad \text{ with } p_j=\frac{\delta_j\rho_j}{1+\delta_j \rho_j}. 
\end{equation}
The approximation suggested in the literature simply consists in replacing $N_j$  by $I_j=1(N_j>0)$, i.e. the number of impacts in $\Delta_j$ is replaced by the presence/absence of an impact in $\Delta_j$. This yields the criterion
\begin{equation} \label{eq:brl.logit}
\BRLlogit (\bbeta)= \sum_{j=1}^J \left\{
I_j \log\left( p_j\right) 
+ (1-I_j) \log\left(1-p_j \right) 
\right\}.   
\end{equation}
In other words, \eqref{eq:brl.logit} is the likelihood of the Bernoulli random variables $I_j$ if one  (wrongly) assumes that $\mathrm{logit}\pr(I_j=1)= \log(\delta_j) + \bbeta^\top \bC_j$. 
Replacing $N_j$ by $I_j$ in~\eqref{eq:clrl} is definitely a rough approximation which is difficult to quantify without any assumption on the underlying spatial point process $\bY$. This approach is also not robust to the Poisson assumption for $\bY$.

\subsection{Bernoulli regression likelihood with complementary \linebreak log-log link (method $\BRLcloglog$)} \label{sec:BRLcloglog}

The previous approximation can be alleviated if we link correctly $\mu_j=\mathrm P(I_j=1)$ to $\bbeta$. A model is required for this. If one assumes that the variables $N_j$ are independent and distributed as Poisson random variables, then standard calculation shows that $g(\mu_j) = \bbeta^\top \bC_j + \log \delta_j$ where $g$ is the log-log complementary function given by $g(t) =\log (-\log(1-t))$ for $t\in (0,1)$. Then, the log-likelihood based on $I_1,\dots,I_J$ is equal to
\begin{align}
\BRLcloglog(\bbeta) &= \sum_{j=1}^J \left\{ I_j \log ( \mu_j ) + (1-I_j)\log(1-\mu_j) \right\} \nonumber\\
&= \sum_{j=1}^J \left\{ I_j \log \left( \frac{\mu_j}{1-\mu_j}\right) + \log(1-\mu_j) \right\}.\label{eq:BRLcloglog}
\end{align}
Therefore, if $\bY$ is a (spatio-temporal) Poisson point process, $\bbeta$ can be indeed (consistently) estimated using a logistic regression with log-log complementary link and  offset term $(\log \delta_j)_j$. However,~\eqref{eq:BRLcloglog} is not a composite likelihood and so is not robust at all to the Poisson assumption. Indeed,  the score of $\BRLcloglog(\beta)$ is equal to 
\[
\BRLcloglog^{(1)} (\bbeta)= \sum_{j=1}^J C_j \log(1-\mu_j)  \left(
 \frac{I_j}{\mu_j}  - 1
\right)
\]
whereby we deduce, by denoting $\tilde \mu_j = \pr(N_j>0)$, that
\[
  \E \left( \BRLcloglog^{(1)}(\bbeta) \right)= \sum_{j=1}^J \bC_j \log(1-\mu_j) \left(\frac{\tilde \mu_j}{\mu_j}-1\right).
\]
For general point processes the assumption that $\tilde \mu_j= \mu_j, \forall j$ does not hold. Hence, $\BRLcloglog^{(1)} (\beta)$ is not, in general, an unbiased estimating function for non-Poisson point processes and thus this procedure should not be recommended if one suspects a significant departure of data from the Poisson assumption.



\subsection{Other approaches} \label{sec:otherapproaches}

As we do not want to model second-order characteristics of the point process $\bY$, the maximum likelihood method is not considered. As far as we know, two other approaches could have been investigated. The first one is the quasi-likelihood method proposed  by~\citet{guan2015quasi}. This method designed for spatial point processes (but which can easily be extended to spatio-temporal point processes) has the merit to be robust to the Poisson assumption and to reduce the variance of the Poisson likelihood estimator. Its computation cost is however important in particular for large datasets. This approach also requires a preliminary estimation of the pair correlation function. Modelling the second-order structure of the spatio-temporal point process $\bY$ is not the topic of this paper. 

The second one, based on a variational type estimating equation, is proposed by~\citet{coeurjolly2014variational}. The main interest of this method is that it does not require any optimization procedure and is very cheap from a computational point of view. However, it requires that covariates are observed at a very fine scale in particular in a neighborhood of the observed points. Such an assumption does not hold for the application considered in this paper.


\section{Zero-deflated subsampled versions of composite likelihoods methods}\label{sec:subsampling}


Given the size of regressions involved in the lightning strikes data ($J\approx 2.8\times 10^6$ and the number of impacts can easily reach $300,000$ for some seasons if one applies any method from Section~\ref{sec:estimation} to $\bY_{s,m}$), the previous methodologies are computationally expensive and can lead to numerical instabilities. A way to alleviate these problems is to use subsampling strategies, which can also yield more robust estimates (see e.g. \citet{humbert2022robust}). In the framework of point processes, subsampling corresponds to a thinning process \citep[see][]{daley2003introduction,moller2003statistical,cronie2023statistical}. In particular, if $\pi$ stands for the retaining probability (i.e. the thinning probability is equal to $1-\pi$) assumed to be constant over space and time for the sake of simplicity, then it is well-known that the intensity $\rho_{\sub}$ of the thinned point process $\bY_{\sub}$ is equal to $\rho_{\sub}(y) = \pi \rho(y)$. Hence if  $\rho(y)= \exp(\bbeta^\top \bC(y))$, all methods described in Section~\ref{sec:estimation} can be applied to estimate $\bbeta$ based on $\bY_\sub$ up to an offset term. All remarks made on parametric methods summarized in Table~\ref{tab:summary} apply in the same way if one considers $\bY_\sub$ instead of $\bY$.  The subsampling procedure could also be repeated and averaged over independent replications as investigated by \citet{cronie2023statistical} for cross-validation techniques. 

However, if we see our dataset as counts of impacts per cell $\Delta_j$ (or presence/absence of an impact per cell), the dataset is highly unbalanced as shown in Figure~\ref{fig:zeroes}. So if one samples cells independently, there is a high probability that the subsample contains only cells with no data point. For such subsamples, parametric methods presented in Section~\ref{sec:estimation} will obviously fail or lead to numerical instabilities. 

This  practical problem is well-known in classification, machine learning tasks and regression problems, see e.g.~\citet{fernandez2018learning,spelmen2018review} and the references therein. In the context of spatial point proceses, \citet[Section 9.10.3]{baddeley2015spatial} have also considered this problem and describe a subsampled version of $\BRLlogit$ (pixel logistic regression) where cells with impacts are more likely to appear in the subsample than empty cells. The goal of this section is to propose extensions of methods described in Section~\ref{sec:estimation} for zero-deflated subsamples, that is for subsamples of $N_j$ or $I_j$ such that cells with no data point are less likely to appear (for instance) in the resulting subsample. A summary of these subsampled versions is proposed in Table~\ref{tab:summary.sub}.

\subsection{Subsampled version of $\BRLlogit$}  \label{sec:BRLlogit.sub}

We formalize and slightly extend \citet[Section 9.10.3]{baddeley2015spatial}. For $\ell=0,1$, let $J_\ell=\{j=1,\dots,J: I_j=\ell\}$ be the set of empty and non-empty cells and let $\mathcal J_\ell$ be a sample from $J_\ell$ with constant inclusion probability $\pi_\ell=\pr (j \in \mathcal J_\ell \mid I_j=\ell)$ and let $\mathcal J=\mathcal J_0\cup \mathcal J_1$. Thus, $\mathcal J$ is a discrete random set with on average $\pi_0 J \pr(I_j=0)$ elements corresponding to empty cells and $\pi_1 J \pr(I_j=1)$ indices corresponding to cells containing at least one data point. The variables $I_j \mid j\in \mathcal J$ are Bernoulli random variables with mean
\begin{equation*}
    \pr(I_j=1 \mid j \in \mathcal J )  = \pr(j \in \mathcal J \mid I_j=1 ) \frac{\pr(I_j=1)}{\pr(j\in \mathcal J)} = \frac{\pi_1}{\pr(j\in \mathcal J)} \pr(I_j=1).
\end{equation*}
Proceeding similarly for $\pr(I_j=0 \mid j \in \mathcal J )$ yields
\begin{equation} \label{eq:pI1surpI0}
    \frac{\pr(I_j=1 \mid j \in \mathcal J ) }{\pr(I_j=0 \mid j \in \mathcal J ) } = \frac{\pi_1}{\pi_0} \, \frac{\pr(I_j=1)}{\pr(I_j=0)}.
\end{equation}
Now, let us keep the spirit of the method $\BRLlogit$. \rev{If}
\begin{equation}
\label{eq:wrong}
\rev{\mathrm{logit} \pr(I_j=1) = \log(\delta_j)+ \bbeta^\top \bC_j,  } \end{equation}
\rev{then,  the likelihood of the $I_j \mid j\in \mathcal J$ can {\it still} be used to estimate $\bbeta$ using a standard logistic regression, but now with offset term $(\log(\delta_j)+\log(\pi_1/\pi_0))_{j\in\mathcal J}$.  As already mentioned in Section~\ref{sec:BRLlogit}, \eqref{eq:wrong} has no reason to hold. However, Equation~\eqref{eq:pI1surpI0} is the cornerstone of Section~\ref{sec:subsampling} that we exploit to provide extension of the methods $\BRLcloglog, \CLRL, \WCLRL$ and $\PL$.}

\subsection{Subsampled version of $\BRLcloglog$} \label{sec:BRLcloglog.sub}

We sample as in the previous section but we now assume that $\bY$ is, as in~Section~\ref{sec:BRLcloglog}, a spatio-temporal Poisson point process, which implies that $\pr(I_j=1)=1-\exp(\delta_j \exp(\bbeta^{\top} \bC_j))$. Using this and~\eqref{eq:pI1surpI0}, we can deduce that 
\begin{equation}
g^{-1} \left( \pr(I_j=1\mid j\in \mathcal J)\right) = \log \delta_j + \bbeta^\top \bC_j    
\end{equation} 
where for any $t\in (0,1)$
\begin{equation} \label{eq:cloglog.sub}
g^{-1}(t) = \log\left( \log \left( 1+ \frac{\pi_0}{\pi_1} \frac{t}{1-t}\right)\right).    
\end{equation}
In other words, given $\mathcal J$ and under the Poisson assumption, the likelihood of the Bernoulli random variables $I_j, j\in \mathcal J$, corresponds to a logistic regression with link function~\eqref{eq:cloglog.sub} and offset term $(\log \delta_j)_{j\in \mathcal J}$. It is worth pointing out that when $\pi_0=\pi_1$, the link function~\eqref{eq:cloglog.sub} reduces to the complementary log-log function. This is why we suggest the name of \texttt{clogclog.sub} for this link function (see Table~\ref{tab:summary.sub}). Similarly to the method $\BRLcloglog$, its subsampled version is in general not robust to the Poisson assumption.

\subsection{Subsampled version of $\CLRL$ and $\WCLRL$} \label{sec:CLRL.sub}

Following Sections~\ref{sec:CLRL} and~\ref{sec:BRLlogit.sub}, we can propose the following subsampled version of $\CLRL$ (the subsampled version of $\WCLRL$ is similar). Let $\mathcal N$ (resp. $\mathcal J$) be a sample of $\{j=1,\dots,N+J: \tilde I_j=1\}$ (resp. $\{j=1,\dots,N+J: \tilde I_j=0\}$) with inclusion probabilities $\pi_1=\pr(j\in \mathcal N \mid \tilde I_j=1)$, $\pi_0=\pr(j\in \mathcal J \mid \tilde I_j=0)$ and let $\mathcal K= \mathcal N\cup \mathcal J$. Then, it is clear from Section~\ref{sec:BRLlogit.sub} that
\begin{equation*}
\mathrm{logit} \; \pr( \tilde I_j=1 \mid j\in \mathcal K)= \log\left( \frac{\pi_1}{\pi_0}\right) + \log(\delta_j) + \mathrm{logit}\; \pr( \tilde I_j=1 ).  
\end{equation*}
A subsampled version of $\CLRL$ is easily deduced: we estimate $\bbeta$ by a logistic regression of
$(\tilde I_j)_{j\in \mathcal K}$ in terms of $(\bC_j)_{j\in \mathcal K}$ with logit link and offset term $(\log \pi_1/\pi_0 + \log \delta_j)_{j\in\mathcal K}$. It is worth pointing out that the subsampled version of $\CLRL$ (and $\WCLRL$) is fundamentally different from the other methods in the sense that we sample according to the values of the $\tilde I_j$, i.e. we sample from the data points with probability $\pi_1$ and independently from the cells with probability $\pi_0$. This subsampling is less natural than the one used for the other methods where we sample from cells with or without impact.

\subsection{Subsampled version of $\PL$} \label{sec:PL.sub}

As seen in Section~\ref{sec:PL}, the method $\PL$ corresponds to a Poisson regression of the random variables $N_j$  with the canonical log link. So a subsampled version of this method can be designed by subsampling these random variables. For any $\ell \in \mathbb N$, let $\mathcal J_\ell$ be a sample from $J_\ell=\{j=1,\dots,J: N_j=\ell\}$ with inclusion probability $\tau_\ell = \pr(j\in \mathcal J_\ell \mid N_j=\ell)$. Given $\mathcal J=\cup_{\ell\ge 0} \mathcal J_\ell$, the $N_j$'s are discrete random variables with (conditional) mean
\begin{align}
\E \left( N_j \mid j \in \mathcal J\right)     &= \sum_{\ell\ge 0} \ell \, \pr(N_j=\ell \mid j \in \mathcal J_\ell)  \nonumber\\
&= \sum_{\ell=0}^\infty \ell \, \pr( j \in  \mathcal J_\ell \mid N_j=\ell ) \frac{\pr(N_j=\ell)}{\pr(j\in \mathcal J_\ell)}  \nonumber\\
&=\sum_{\ell=0}^\infty \ell \, \pr( j \in \mathcal J_\ell \mid N_j=\ell ) \frac{\pr(N_j=\ell)}{\pr(j\in \mathcal J)} \nonumber\\
&= \sum_{\ell>0} \ell \, \pr(N_j=\ell) \frac{\tau_\ell}{\pr(j\in \mathcal J)}. \label{eq:Nj.sub}
\end{align}
Assuming $\tau_0=\pi_0$ and $\tau_\ell=\pi_1$ for any $\ell\ge 1$, which implies we sample according to the values of $I_j$ as in Sections~\ref{sec:BRLlogit.sub}-\ref{sec:BRLcloglog.sub}, leads to 
\begin{equation*}
\E (N_j \mid  j\in \mathcal J) = \frac{\pi_1}{\pr(j \in \mathcal J)} \E (N_j) 
\end{equation*}
and then to
\begin{equation*}
\log    \E (N_j \mid  j\in \mathcal J) = \log\left(\frac{\pi_1}{\pr(j \in \mathcal J)}\right) + \log(\delta_j) + \bbeta^\top \bC_j.
\end{equation*}
The problem is that $\pr(j\in \mathcal J)$ is unknown and depends on the parameter vector $\bbeta$. The likelihood of the $N_j \mid j\in\mathcal J$ is therefore no more a Poisson regression. We suggest to estimate $\pr(j \in \mathcal J)$ by $\pi_0\mathbf{1}(j\in \mathcal J_0)+\pi_1\mathbf{1}(j\in \mathcal J_1)$. This estimation leads to
\begin{equation*}
\log    \E (N_j \mid  j\in \mathcal J) = o_j +  \beta^\top C_j \qquad \text{ with } o_j = \log\left(\frac{\pi_1}{\pi_0} \mathbf 1(j\in J_0) \right) + \log(\delta_j)
\end{equation*}
which can now be implemented by a Poisson regression with log link and offset term $(o_j)_{j\in \mathcal J}$.



\subsection{A brief simulation study, remarks and possible extensions} \label{sec:sim}

In this section, we propose a short simulation study to underline the interest and efficiency of zero-deflated subsampling.
We consider a spatio-temporal Poisson point process with intensity model $\rho(y)=\exp(\bbeta^\top \bC(y))$ where $\bbeta \in \mathbb R^3$ ($n_C=3$) with $\beta_2=\beta_3=1$. The first covariate is 1 (thus $\beta_1$ is an intercept term) and the covariate functions $C_2(y)$ and $C_3(y)$ are discretized such that $C_{2j},C_{3j}$ are realizations of standard normal random variables for $j=1,\dots,J$. Thus, the $N_j$'s are Poisson random variables with mean $\mu_j=\exp(\bbeta^\top \bC_j)$ and we  adjust $\beta_1$ such that on average on the sample with size $J$, we have 50\%, 90\%, 99\% or 99.9\% of 'empty cells'. 

\begin{figure}[htbp] 
\centering
\includegraphics[width=.8\textwidth]{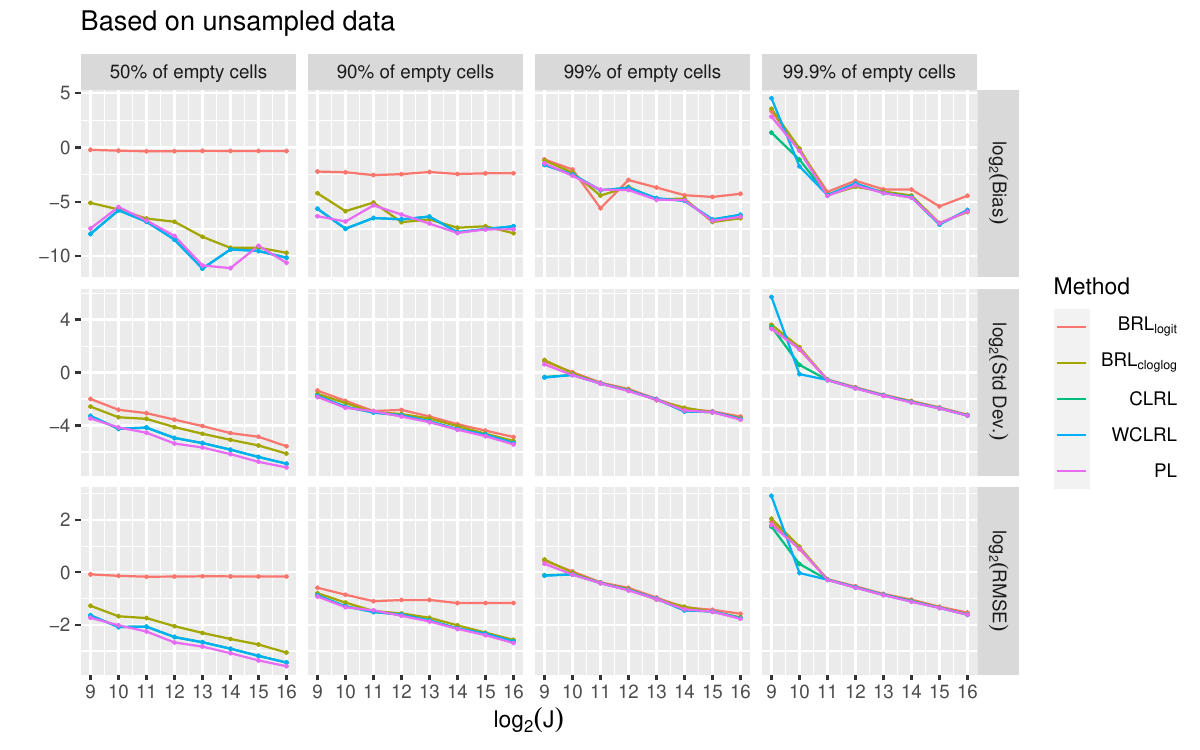}
\includegraphics[width=.8\textwidth]{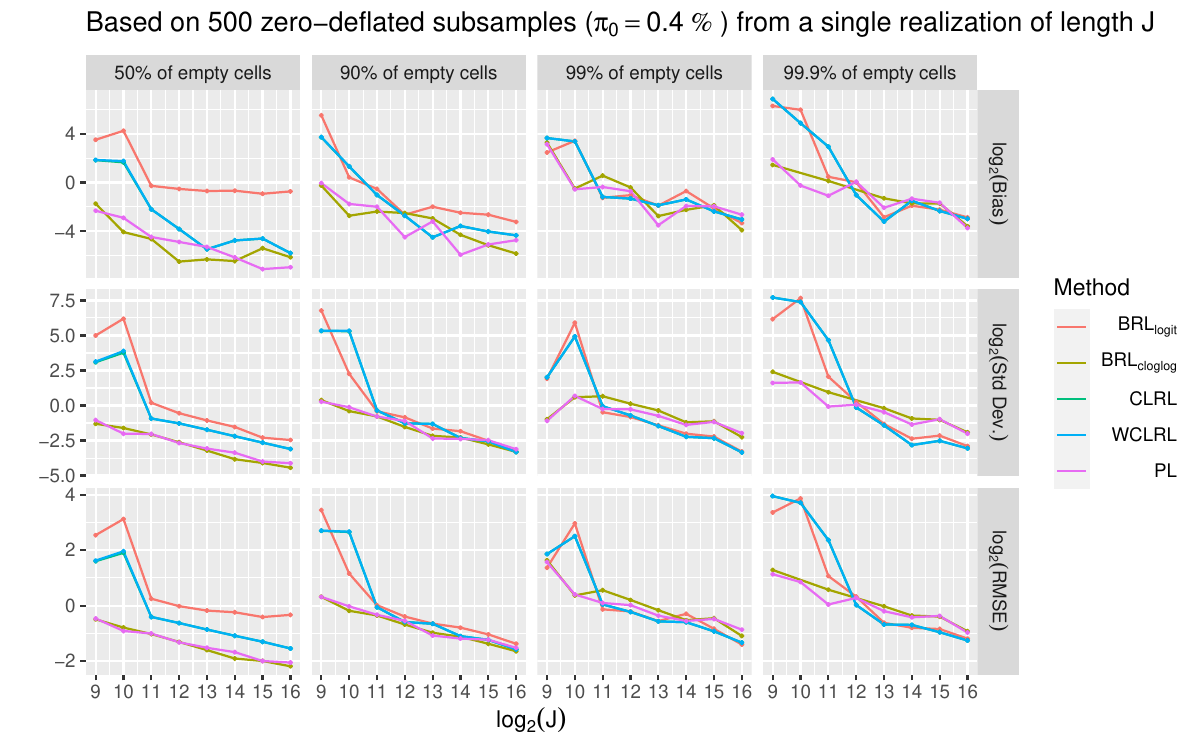}
\caption{Logarithms (in base 2) of empirical (average absolute) biases, standard deviation and root-mean squared errors for estimates of $\bbeta=(\beta_1,1,1)$  in terms of $\log_2(J)$ based on $B=500$ replications of the simulation model described in Section~\ref{sec:sim} yielding 50\%, 90\%,  99\% or 99.9\%. (Top) Data are unsampled (Bottom) From one simulation of length $J$, we subsample 500 replications with $\pi_1=1$ and $\pi_0=0.4\%$.}  \label{fig:simulation}
\end{figure}

Figure~\ref{fig:simulation} (top) constitutes the baseline. We generate $B=500$ replications of the previous model and estimate $\bbeta$ using methods described in Section~\ref{sec:estimation} (no subsampling). We depict the logarithms of the empirical (average absolute) biases, standard deviations and root mean squared errors (RMSE) in term of $\log_2(J)$ for $J=2^k$ and  $k=9,\dots,16$. As mentioned in Section~\ref{sec:estimation}, the method $\BRLlogit$ is the only method which is based on a biased estimating equation due to the misspecification of the link function. That is why this method leads to non negligible bias. All other methods lead to asymptotically unbiased estimates. We note first that the larger the number of empty cells, the larger the bias and second that the difference with the method $\BRLlogit$ is less obvious when the proportion of empty cells reaches 99.9\%. In terms of variance, we first observe the linear trend which means that the variance of estimates seems to decrease as a power $J$ and this for any configuration of 'empty cells'. The best method seems to be the Poisson likelihood one (method $\PL$). Again the variances are higher and the differences between all methods less important when the proportion of empty cells increases. 

To evaluate methods described in Section~\ref{sec:subsampling} which are based on subsampling strategies, we  start with a single simulation with length $J=2^k (k=9,\dots,16)$ and estimate $\bbeta$ based on $B=500$ subsamples. We set $\pi_1=1$ and $\pi_0=2^{-2}=25\%$, then $\pi_0=2^{-5}=3.1\%$ (Figure~\ref{fig:simulation2}) and $\pi_0=2^{-8}=0.4\%$ (Figure~\ref{fig:simulation} (bottom). Thus, for these simulations, for each value of $J$ and $\pi_0$, estimates are based on data with length on average $J(p+\pi_0(1-p))$ where $p$ if the proportion of non-empty cells. The behaviors of empirical biases, standard deviations and thus root mean squared errors are more erratic than in the situation where $\pi_0=1$ (no subsampling). Also the values of the different scores are higher than the baseline situation but the comments on the different methods still apply. In particular the root mean squared errors tend to 0 with $J$ (except for the method $\BRLlogit$) for any value of $1-p$ (the proportion of empty cells). We have presented here results for a very small value of $\pi_0$ (more moderate situations are presented in Figure~\ref{fig:simulation2}). This intrinsically shows the efficiency of zero-deflated procedures presented in Section~\ref{sec:subsampling} including the definition and estimations of offset terms. 

This section convinces us of the interest of zero-deflated subsampling type procedures. Overall, when the proportion of empty cells is upto 90\% or for moderate sample sizes and higher value of $1-p$, the Poisson likelihood could be recommended. In extreme situations and for large sample sizes, it is harder to distinguish the different methods. 
Finally, as expected there are no differences between the methods $\CLRL$ and $\WCLRL$. 
Only the last one is used in Section~\ref{sec:backDataset} as it is faster than the other one.


\renewcommand*{\arraystretch}{1.6}
\begin{landscape}
\begin{table}[htbp]
\begin{tabular}{p{.18\textwidth}p{.2\textwidth}p{.22\textwidth}p{.22\textwidth}p{.22\textwidth}p{.2\textwidth}}
\hline 
Method& ${\BRLlogit}$& $\BRLcloglog$& $\CLRL$& $\WCLRL$& $\PL$\\
&(see Section~\ref{sec:BRLlogit})&(see Section~\ref{sec:BRLcloglog})&
(see Section~\ref{sec:CLRL})&(see Section~\ref{sec:CLRL})&(see Section~\ref{sec:PL})\\
\hline \hline
GLM family & \texttt{binomial}&\texttt{binomial} & \texttt{binomial} &\texttt{binomial}& \texttt{poisson} \\
\hline
Link & \texttt{logit}&\texttt{cloglog}&\texttt{logit}&\texttt{logit}&\texttt{log}\\
\hline
R formula & {\texttt{I\~{}C}} &\texttt{I\~{}C} &  
\texttt{Icheck\~{}Ccheck} &\texttt{Itilde\~{}Ctilde}, \texttt{w=wtilde} &
\texttt{N\~{}C}\\
\hline
Offset term & $(\log \delta_j)_j$ & $(\log \delta_j)_j$ & $(\log \delta_j)_j$ & $(\log \delta_j)_j$ & $(\log \delta_j)_j$ \\
\hline
Size of the response vector & $J$ &$J$ &$N+J$ &$\#J_1+J$ \\ 
\hline
Unbiased score& $\times$ & $\checkmark$ (if Poisson) & $\checkmark$ & $\checkmark$ & $\checkmark$ \\
\hline
Robustness& $\times$ & $\times$ &$\checkmark$&$\checkmark$&$\checkmark$  \\
\hline
Variance & +/- & +/- & + & + & ++ \\
\hline
\end{tabular}
\caption{Summary of subsampled versions of parametric methods used to estimate an exponential family model with piecewise constant covariates; \texttt{I} (resp. \texttt{Icheck,Itilde,N}) stands for the vector of the $I_j$ (resp. $\check I_j$, $\tilde I_j$, $N_j$); \texttt{C} (resp. \texttt{Ccheck,Ctilde}) stands for the matrix of constant covariates; $\delta_j$ is the volume of the spatio-temporal cell $\Delta_j$. \rev{A $\checkmark$ (resp. $\times$) means that the statement is correct (resp. wrong). A $+$ means that the variance is shown to converge to 0 when the amount of data increases under general models. The ++ means that the variance is the lowest among considered methods. A +/- means that the variance decreases only for some specific situations.}}
\label{tab:summary}
\end{table}
\end{landscape}

\begin{landscape}
\begin{table}[htbp] 
\begin{tabular}{p{.18\textwidth}p{.2\textwidth}p{.22\textwidth}p{.22\textwidth}p{.22\textwidth}p{.2\textwidth}}
\hline 
Subsampled version of& ${\BRLlogit}$& $\BRLcloglog$& $\CLRL$& $\WCLRL$& $\PL$\\
&(see Section~\ref{sec:BRLlogit.sub})&(see Section~\ref{sec:BRLcloglog.sub})&
(see Section~\ref{sec:CLRL.sub})&(see Section~\ref{sec:CLRL.sub})&(see Section~\ref{sec:PL.sub})\\
\hline \hline
GLM family & \texttt{binomial}&\texttt{binomial} & \texttt{binomial} &\texttt{binomial}& \texttt{poisson}  \\
\hline
Link & \texttt{logit}&\texttt{cloglog.sub}&\texttt{logit}&\texttt{logit}&\texttt{log}\\
\hline
R formula & {\texttt{I.sub\~{}} \texttt{C.sub}} &\texttt{I.sub\~{}} \texttt{C.sub} &   
\texttt{Icheck.sub\~{}} \texttt{Ccheck.sub} &\texttt{Itilde.sub\~{}} \texttt{Ctilde.sub}, \texttt{w=wtilde.sub} & 
\texttt{N.sub\~{}} \texttt{C.sub}\\
\hline
Offset term 
& $\big( \log\frac{\pi_1}{\pi_0}$ $+\log \delta_j\big)_j$ 
& $(\log \delta_j)_j$ 
& $\big( \log\frac{\pi_1}{\pi_0}$ $+\log \delta_j\big)_j$ 
& $\big( \log\frac{\pi_1}{\pi_0}$ $+\log \delta_j\big)_j$ 
& $\big(\log\big(\frac{\pi_1}{\pi_0} \big)$ $\mathbf 1(j\in \mathcal J_0)$ 
$+\log \left(  \delta_j\right)\big)_j$
\\ 
\hline
Size of the response vector 
& $\#\mathcal J_0+\#\mathcal J_1$ 
& $\#\mathcal J_0+\#\mathcal J_1$ 
& $\#\mathcal N+\#\mathcal J$
& $\#\mathcal N_1+\#\mathcal J$
& $\#\mathcal J_0+\#\mathcal J_1$ 
\\
\hline
{\small Unbiased score given subsample}& $\times$ & $\checkmark$ (if Poisson) & $\checkmark$ & $\checkmark$ & $\approx \checkmark$ \\ \hline
Robustness& $\times$ & $\times$ &$\checkmark$&$\checkmark$&$\checkmark$  \\
\hline
\end{tabular}
\caption{Summary of characteristics of parametric methods used to estimate an exponential family model with piecewise constant covariates; \texttt{I.sub} (resp. \texttt{Itilde.sub,Icheck.sub,N.sub,}), stands for a (dependent) subsample of the $I_j$  (resp. $\tilde I_j$, $\check I_j$, $N_j$); \texttt{C.sub} (resp. \texttt{Ccheck.sub,Ctilde.sub}) stands for the subsampled matrix of constant covariates; $\delta_j$ is the volume of the spatio-temporal cell $\Delta_j$; (see Section~\ref{sec:subsampling} for a definition $\pi_0$, $\pi_1$, $\mathcal J_0$, $\mathcal J_1$, $\mathcal N$,  $\mathcal J$ and $\mathcal N_1$ according to each method). \rev{A $\checkmark$ (resp. $\times$) means that the statement is correct (resp. wrong). The symbol $\approx \checkmark$ means that we have observed the unbiasedness upon simulations but this does not follow from a theoretical result.} }
\label{tab:summary.sub}
\end{table}
\end{landscape}

\section{Back to the lightning strikes dataset} \label{sec:backDataset}

\newcommand{\learn}{\text{learn}}
\newcommand{\test}{\text{test}}

We return to the dataset described in Section~\ref{sec:desc}, and model it as a spatio-temporal point process with intensity given by an exponential family with discretized spatial and spatio-temporal covariates (see details in  Section~\ref{sec:background}). In the following, 
we decompose the spatio-temporal point process $\bY$ into $\bY^{\learn}$ and $\bY^{\test}$  as
\begin{align*}
\bY^{\learn} &= \{(x,t) \in W \times \mathcal T^{\learn}\} 
\text{ with } \mathcal T^{\learn} =\{t \in \R^+: \text{year} \in \{2013,2014\} \} \\
\bY^{\test} &= \{(x,t) \in W \times \mathcal T^{\test}\} 
\text{ with } \mathcal T^{\test} =\{t \in \R^+: \text{year} \in \{2015\} \} \; . 
\end{align*}
The goal of this section is to estimate the spatio-temporal parametric intensity function using $\bY^\learn$ as specified in Section~\ref{sec:background}, and use the fitted model to predict the spatio-temporal intensity of $\bY^{\test}$, so in 2015. More precisely, for any $s,m\in \mathcal S\times \mathcal M$ (ie for any season and 6-hour period of the day), we predict the spatio-temporal intensity of $\bY_{s,m}^\test$ by learning the one of $\bY^{\learn}_{s,m}$. Let $W\times \mathcal T_{s,m}^{\test}= \bigcup_{j\in \mathcal J_{s,m}^\test} \Delta_j$. Thus $\mathcal J_{s,m}^{\test}$   is  the set of indices of cells corresponding to $s,m$ in the test dataset and we intend, in particular, to compare the variables $N_{j,s,m}$ and $\hat N_{j,s,m}$ for $j\in \mathcal J_{s,m}^{\test}$,  where
\begin{align*}
N_{j,s,m}= N(\Delta_{j,s,m}) = \sum_{y \in \bY^{\test}_{s,m}} \mathbf 1(y \in \Delta_{j,s,m}) \\    
\hat N_{j,s,m}= \int_{\Delta_{j,s,m}} \exp( \hat \bbeta_{s,m}^\top \bC(y) ) \dd y   =  \exp( \hat \bbeta_{s,m}^\top \bC_j )
\end{align*}
since the covariates are constant over $\Delta_{j,s,m}$ and since we assume without loss of generality that $\delta_{j,s,m}=|\Delta_{j,s,m}|=1$. In what precedes, one has of course $\hat \bbeta_{s,m} = \hat \bbeta_{s,m}(\bY^{\learn})$. We can also decide to compare $N_{j,s}$ and $\hat N_{j,s}$, ie observations and predictions averaged over the 6-hour periods, or $N_j$ and $\hat N_j$, ie observations and predictions averaged over 2015. 

Evaluating the quality of predictions is a tricky task for which several approaches exist. Section~\ref{sec:criteria} discusses the criteria we focused on, while Section~\ref{sec:res} provides empirical results on the lightning strikes dataset.

We end this section by emphasizing that we use the term 'prediction' in the machine learning way. Our goal is to characterize globally the intensity of lightning strikes by spatial and spatio-temporal covariates and we use training and test sets to compare the methodology proposed in this paper. We do not mean 'prediction' as a short-term forecast. 

\subsection{Criteria used to measure the quality of predictions} \label{sec:criteria}

To evaluate the ability of different methods to predict the presence/absence of lightning strikes, we consider for any $s,m$ and any $j\in \mathcal J_{s,m}^{\test}$, 
$I_{j,s,m}=\mathbf 1(N_{j,s,m}>0)$ and $\hat I_{j,s,m}=\mathbf 1(\hat N_{j,s,m}>0)$ and compute standard metrics such as areas under ROC Curve and Precision-Recall Curve, denoted respectively by $\text{AUC}_{s,m}$ and $\text{PPV}_{s,m}$; see e.g.  
\citet[Section 5.7.2]{murphy2012}.

To evaluate the ability of predicting higher values of number of impacts, we suggest the metric $\text{WW}_{s,m}$ (for weighted Wasserstein). This metric, justified later, is defined as
\begin{equation}
\label{eq:WW}
\text{WW}_{s,m} = \omega \int_0^1 \left| L_{p,s,m} - R_{p,s,m}\right| \dd p
\end{equation}
with $\omega= f(N_{s,m}/\hat N_{s,m})$ 
where $N_{s,m}$ and $\hat N_{s,m}$ denote the total number on $ \mathcal J_{s,m}^{\test}$ of observed and predicted lightning strikes in 2015's season $s$ and 6-hour period of the day $m$, and 
$f(x) = x \mathbf 1(x \geq 1) + x^{-1} \mathbf 1(x < 1)$. 
 Using the function $\tilde j: \R^d\times \R^+ \to \{1,\dots,J\}$ given by $\tilde j(y)=\sum_{s,m} \sum_{j\in \mathcal J_{s,m}^{\test}} j \mathbf 1(y\in \Delta_{j,s,m})$ which identifies the spatio-temporal cell an impact at location $y$ belongs to, we define for $\bar \nu_{s,m}$, an upper-bound of $\max_{j\in \mathcal J_{s,m}^{\test}} \hat N_{j,s,m}$
\begin{align}
L_{p,s,m} &= \frac{1}{N_{s,m}}\sum_{j \in \mathcal J_{s,m}^{\test}} N_{j,s,m}\mathbf 1(\hat N_{j,s,m}\le p \bar \nu_{s,m}) \nonumber\\
&= \frac{1}{N_{s,m}}\sum_{y \in \bY^{\test}_{s,m} } \mathbf 1(\hat N_{\tilde j(y),s,m}\le p \bar \nu_{s,m}) \label{eq:L}\\
R_{p,s,m} &= \frac{1}{\hat N_{s,m}} \sum_{j\in \mathcal J_{s,m}^{\test}} \hat N_{j,s,m}\mathbf 1(\hat N_{j,s,m}\le p \bar \nu_{s,m}) \nonumber \\
&= \frac1{\hat N_{s,m}} \int_{W\times \mathcal T^\test_{s,m}} \mathbf 1(\hat N_{j,s,m}\le p \bar \nu_{s,m})\rho(y;\hat \bbeta_{s,m}) \dd y. \nonumber
\end{align}
As a first comment, let $\delta_p= N_{s,m}L_{p,s,m}- \hat N_{s,m}R_{p,s,m}$. It is worth pointing out that
\[
{\delta_p =\sum_{y\in \bY_{s,m}^\test} h(y;\bY^\test)-\int_{W\times \mathcal T_{s,m}^\test}h(y;\bY^\test)\rho(y;\hat\bbeta_{s,m}(\bY^\learn))\dd y}
\]
where $h(y;\bY^\test)=\mathbf 1(\hat N_{\tilde j(y),s,m}\le p\bar\nu_{s,m})$. 
Such a random variable is a bivariate residual functional for point processes. More precisely if $\rho(\cdot;\hat \bbeta)=\rho(\cdot;\bbeta)$ (ie the true intensity), $\delta_p$ would be a bivariate innovation functional as defined by~\cite{cronie2023statistical}, proved to be centered under the true model. The criterion~\eqref{eq:WW} can therefore be viewed as a self-normalized weighted combination of generalized residuals.

But we can actually say more about the intuition of this criterion. By Campbell theorem, we see that $R_{p,s,m}$ is close to 
\begin{equation}\label{eq:Rcheck}
\widecheck R_{p,s,m} = \frac1{\hat N_{s,m}} \sum_{y\in \widecheck \bY_{s,m}^\test}  \mathbf 1(\hat N_{\tilde{j}(y),s,m}\le p \bar \nu_{s,m})
\end{equation} 
where $\widecheck \bY^\test_{s,m}$ is a spatio-temporal point process with $\hat N_{p,s,m}$ events and intensity function $\rho(\cdot ;\hat \bbeta_{s,m}(\bY^\learn))$. Now given $N_{s,m}$ (resp. given $\hat N_{s,m}$), we observe that~\eqref{eq:L} (resp.~\eqref{eq:Rcheck}) is the empirical cumulative distribution function evaluated at $p$  based on the sample $\{\hat N_{\tilde{j}(y),s,m}/\bar \nu_{s,m}, y\in \bY^\test_{s,m}\}$ (resp. $\{\hat N_{\tilde{j}(y),s,m}/\bar \nu_{s,m}, y\in \check\bY_{s,m}^\test\}$). Let $\hat F_{s,m}^\test$ and $\check F_{s,m}^\test$ be these two empirical cdf, we deduce that 
\[
\int_0^1 \left| L_{p,s,m} - R_{p,s,m}\right| \dd p \approx
\int_{0}^1 |\hat F_{s,m}^\test (p) - \widecheck F_{s,m}^\test(p)| \dd p
\]
which, given $N_{s,m}$ and $\hat N_{s,m}$ is the standard Wasserstein distance between the two empirical distributions. Finally, the coefficient $\omega=f(N_{s,m}/\hat N_{s,m})$ in~\eqref{eq:WW} is aimed to take into account the total mass of $\bY_{s,m}^\test$ and $\widecheck \bY_{s,m}^\test$. All these intuitions lead us to propose the notation $\text{WW}_{s,m}$ for weighted Wasserstein for the criterion defined by~\eqref{eq:WW}. For several years, the Wasserstein distance has been recognized as a robust optimal transport-based metric with numerous applications in classification and learning algorithms. See, for example, Frogner et al. (2015) \citet{frogner2015learning} or Arjovsky et al. (2017) \citet{arjovsky2017wasserstein} for more insights.

To summarize, on the one hand the
areas under ROC Curve and Precision-Recall Curve, $\text{AUC}_{s,m}$ and $\text{PPV}_{s,m}$ are used to evaluate the ability of different methods to predict the presence/absence of lightning strikes, and on the other hand,
the $\text{WW}_{s,m}$ index is used to evaluate the ability of predicting higher values of number of impacts.

\subsection{Results} \label{sec:res}

As explained in the beginning of Section~\ref{sec:backDataset}, we now estimate the spatio-temporal parametric intensity model from years 2013-2014 via  $\bY^\learn$, and then predict the spatio-temporal intensity of $\bY^{\test}$ (for 2015). This is done by fixing a season $s$ and a 6-hour period of the day $m$. For the sake of brevity, we show the results for two pairs of $(s,m)$ only, namely (summer, 12:00-18:00) and (winter, 00:00-06:00), which respectively correspond to cases with the most and the least proportion of non empty cells among all cells. 
The competitors defined in Section~\ref{sec:estimation} are combined with the subsampling strategies detailed in Section~\ref{sec:subsampling}. All the combinations are tested with the whole sample ($\pi_1 = 100\%, \pi_0 =100\%$), and for different subsampling rates $\pi_1 = 1,\pi_0 \in \{10\%,1\%,0.1\%\}$,  each time aggregating either on $1,3$ of $10$ bags (ie subsamples). 

Tables~\ref{tab:summer12am}-\ref{tab:winter00am}
summarize the results obtained. We report the values of prediction criteria PPV, AUC and WW detailed in Section~\ref{sec:criteria}, as well as the calculation time per bag. \rev{Implementation is realized within the \texttt{R} software \citep{logicielR} and in particular using the package \texttt{glmnet} \citep{glmnet}. Since the intensity function is estimated separately for each season and period of day, the computations are parallelized on a 20-core cluster. The overall computational time is essentially due to the estimation procedure while the cost of the prediction step mainly comes from the storage of predicted values for each season, period of day and pixel grid. }

From Tables~\ref{tab:summer12am}-\ref{tab:winter00am}, a few comments can be formulated:
first of all, whatever the criterion, it seems useless to choose more that one bag; second, note that AUC does not appear so informative, since all methods and subsampling rates achieve the same kind of performance. This is also the case for the  PPV criterion in winter 00:00-06:00 (Table~\ref{tab:winter00am}). 
Note also that PPV is quite small, in particular for winter (see Table~\ref{tab:winter00am}). This confirms that the prediction of lightning storms is a very difficult task. However, this should be compared to the PPV  achieved by the random classifier, which is the proportion of non-empty cells for these pairs of (season, moment) ($0.0154$ for (summer, 12:00-18:00) and $0.0004$ for  (winter, 00:00-06:00)). Two methods ($\PL$ and $\BRLcloglog$) stand out for their stability concerning the subsampling rate, as demonstrated by the behaviors of PPV and WW in Table~\ref{tab:summer12am}. Additionally, in the absence of subsampling ($\pi_0 = 1$), PPV lacks discrimination. However, in terms of WW, $\PL$ and $\WCLRL$ slightly outperform the other methods. Notably, subsampling significantly impacts computational time reduction (by at least a factor of 40) without compromising results.

\begin{figure}[htbp]
\centering
\includegraphics[width=\textwidth]{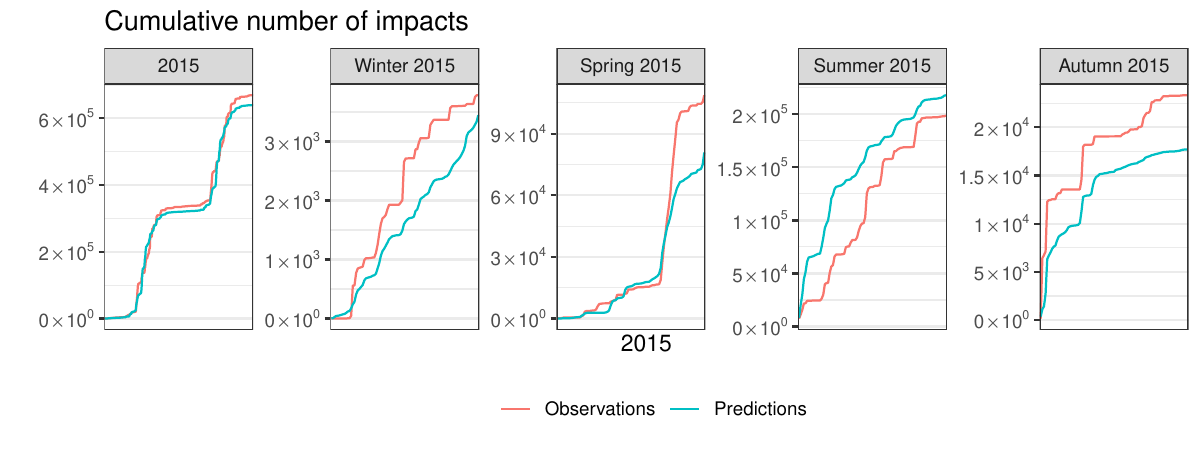} 
\includegraphics[width=\textwidth]{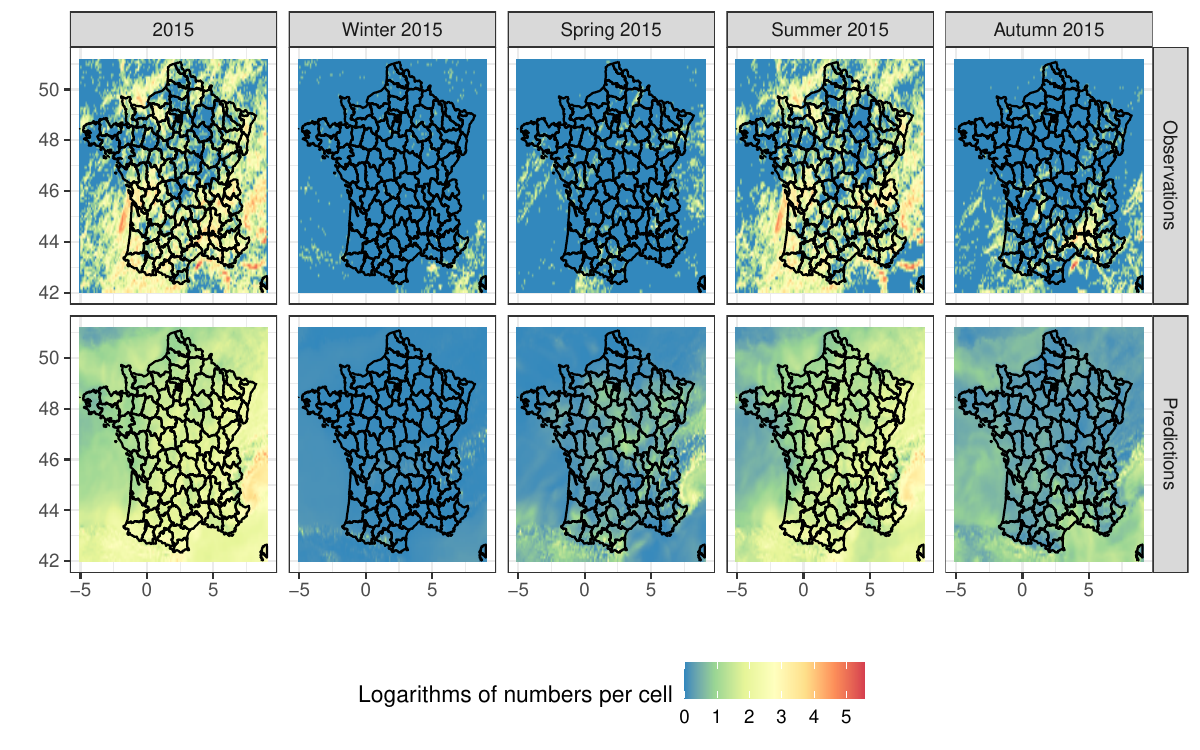}
\caption{Observed and predicted cumulative number of impacts (top) and maps of logarithms of numbers per cell (bottom). Results are presented for each season in 2015 (thus aggregated over periods of days) and globally for  2015.}
\label{fig:pred}
\end{figure}

In a nutshell, the Poisson method $\PL$  with $\pi_0 = 0.1\%, N_{bag} = 1$ appears as the best trade off between prediction quality and computation cost. This method and parameters are used in Figure~\ref{fig:pred} to illustrate the quality of predictions over time and space in 2015. We report observed and predicted cumulative number of impacts  and maps of intensities  over a season of 2015 and over 2015. The cumulative number of impacts are quite well predicted even if some departures can be observed in the end of spring and summer. From a spatial point of view, it is interesting that the fitted model is able to reproduce most of activity phenomena close to the Channel (North West), the French Alps (South East) and Pyrenees. There are however a few areas which are not well captured: the summer activity in the area of Bordeaux (South West), the Winter and summer activity north to Corsica island.


\renewcommand*{\arraystretch}{1.4}
\newcommand{\Nb}{N_{\text{bags}}}

\begin{table}[htbp]
\centering
\begin{tabular}{l|llll}
\hline
& \multicolumn{4}{c}{Method}\\
& $\BRLlogit$ & $\BRLcloglog$ & $\WCLRL$ & $\PL$\\
\hline \hline
$\text{AUC}_{s,m}$ \\
$\pi_0=100\%, \Nb=1$& 89.7 & 89.6 & 89.9 & 89.8  \\
$\pi_0=10\%, \Nb=1,3,10$& 89.7 & 89.6 & 89.9 & 89.8  \\
$\pi_0=1\%, \Nb=1,3,10$& 89.7 & 89.6-89.5 & 89.9 & 89.8  \\
$\pi_0=0.1\%, \Nb=1,3,10$& 89.1-89.3 & 89.8 & 89.1 & 89.7-89.8  \\
\hline
$\text{PPV}_{s,m}$ \\
$\pi_0=100\%, \Nb=1$&12.9&13 & 12.2& 12.5\\
$\pi_0=10\%, \Nb=1,3,10$ & 12.3 & 13.2 & 11.3 & 12.5 \\
$\pi_0=1\%, \Nb=1,3,10$&  11.1-11 & 13.4-13.5 & 9.8-9.9 & 12.8-12.5\\
$\pi_0=0.1\%, \Nb=1,3,10$& 9.1-9.4 & 12.9-12.8 & 8.7 & 11.9-12.6 \\
\hline
$\text{WW}_{s,m}$ \\
$\pi_0=100\%, \Nb=1$& 8.1 & 6.4 & 3.5 & 3.2\\
$\pi_0=10\%, \Nb=1,3,10$ &9.4-9.3 & 5.8-5.9 & 4.9 & 3.2  \\
$\pi_0=1\%, \Nb=1,3,10$& 11.9-12.1 & 12-12.2 & 9-8.6 & 3.2-3.1\\
$\pi_0=0.1\%, \Nb=1,3,10$& 13.2-12.8 & 5.2-5 & 19.7-18.5 & 4.5-3.1 \\
\hline
Time (in sec.) per bag\\
$\pi_0=100\%$& 58 & 98 & 47 & 46 \\ 
$\pi_0=10\%$ &  6 & 23 & 5 & 6 \\
$\pi_0=1\%$&   1 & 6 & 2 & 2  \\
$\pi_0=0.1\%$&  1 & 3 & 1 & 1\\
\hline
\end{tabular}
\caption{Evaluation of prediction metrics ($\text{AUC}_{s,m}$, $\text{PPV}_{s,m}$ and $\text{WW}_{s,m}$) and computational time for subsampled composite likelihood methods for different values of $\pi_0$ and $N_{\text{bags}}$ ($\pi_1=1$). Predictions are done for the season $s$=summer  and the time period $m$=12:00-18:00. Results for $\text{AUC}_{s,m}$, $\text{PPV}_{s,m}$ and $\text{WW}_{s,m}$ have been multiplied by 100. Most of criteria are similar for different values of $\Nb$. When these numbers are different, we provide the range of values (the first, resp. the second, value corresponding to $\Nb=1$, resp. $\Nb=10$).}\label{tab:summer12am}
\end{table}

\begin{table}[htbp]
\centering
\begin{tabular}{l|llll}
\hline
& \multicolumn{4}{c}{Method}\\
& $\BRLlogit$ & $\BRLcloglog$ & $\WCLRL$ & $\PL$\\
\hline \hline
$\text{AUC}_{s,m}$ \\
$\pi_0=100\%, \Nb=1$& 93.1 & 93.2 & 93.3 & 93.3\\
$\pi_0=10\%, \Nb=1,3,10$& 93.1 & 93.2 & 93.3 & 93.3 \\
$\pi_0=1\%, \Nb=1,3,10$& 93 & 93.2-93.1 & 93.2 & 93.2-93.3  \\
$\pi_0=0.1\%, \Nb=1,3,10$& 92-92.4 & 92.7-93.4 & 92.9 & 92.9-93.4  \\
\hline
$\text{PPV}_{s,m}$ \\
$\pi_0=100\%, \Nb=1$& 2 & 2.1 & 2.3 & 2.3 \\
$\pi_0=10\%, \Nb=1,3,10$ & 2 & 2.1 & 2.3 & 2.3\\ 
$\pi_0=1\%, \Nb=1,3,10$&  2 & 2.1 & 2.2 & 2.2-2.3\\
$\pi_0=0.1\%, \Nb=1,3,10$& 1.3-1.6 & 2-2.1 & 1.8-1.9 & 2-2.3  \\
\hline
$\text{PPV}_{s,m}$ \\
$\pi_0=100\%, \Nb=1$&  58.8 & 58.8 & 42.4 & 42.4 \\
$\pi_0=10\%, \Nb=1,3,10$ &57.5-58.9 & 57.4-58.4 & 43.5-42.7 & 42.7-41.4    \\
$\pi_0=1\%, \Nb=1,3,10$& 61.8-65.6 & 51.3-53.3 & 48-45.2 & 39.7-42.6\\
$\pi_0=0.1\%, \Nb=1,3,10$& 40.3-53.9 & 38.8-47.2 & 40.2-45& 25.8-41.9\\
\hline
Time (in sec.) per bag\\
$\pi_0=100\%$& 86 & 127 & 85 & 76 \\ 
$\pi_0=10\%$ & 7 & 11 & 7 & 6 \\
$\pi_0=1\%$&0.9 & 2 & 0.6 & 0.6   \\
$\pi_0=0.1\%$& 0.3& 0.5 & 0.1 & 0.1 \\
\hline
\end{tabular}
\caption{Evaluation of prediction metrics ($\text{AUC}_{s,m}$, $\text{PPV}_{s,m}$ and $\text{WW}_{s,m}$) and computational time for subsampled composite likelihood methods for different values of $\pi_0$ and $N_{\text{bags}}$ ($\pi_1=1$). Predictions are done for the season $s$=winter  and the time period $m$=0:00-6:00. Results for $\text{AUC}_{s,m}$, $\text{PPV}_{s,m}$ and $\text{WW}_{s,m}$ have been multiplied by 100. Most of criteria are similar for different values of $\Nb$. When these numbers are different, we provide the range of values (the first, resp. the second, value corresponding to $\Nb=1$, resp. $\Nb=10$).}
\label{tab:winter00am}
\end{table}

\section{Conclusion} \label{sec:conclusion}

In this paper, we have proposed zero-deflated subsampled versions of standard composite likelihoods methods to estimate exponential family intensity models for spatio-temporal point processes. We have in particular focused on the situation where the covariates are piecewise constant taking the opportunity to review standard composite likelihood methods in this situation. We have theoretically justified these subsampled extensions. The methods perform well on simulations, at least for spatio-temporal Poisson point processes. When they are applied to the lightning strikes dataset, we have concluded that the subsampled Poisson likelihood provides the best compromise in terms of stability, prediction errors and computational time.

We believe this work leads to several perspectives. It would be first interesting to study more deeply these procedures. For instance, can we still be ensured that the estimator derived by these procedures is consistent and asymptotically normal? Extending these procedures for non piecewise constant covariates. Investigating inhomogeneous subsampling, i.e. for instance letting $\pi_0$ and $\pi_1$ depend on the cell $\Delta_j$ are also interesting perspective. \rev{Finally, there is now a large literature on feature selection for point processes in particular using regularization techniques, e.g.~\citet{choiruddin:coeurjolly:letue2018}. Understanding how these techniques can be combined with the zero-deflated subsampling procedure proposed in this paper is of particular relevance.}

\rev{Regarding the dataset application, this paper has explored the spatio-temporal first-order trend of lightning strikes by assuming nothing on the second-order structure. This is standard when one anlyzes a point pattern, e.g.~\citet{waagepetersen2009two}. Based on the present data analysis, modelling second-order characteristics of the process or even better modelling the distribution of $\bY$ for instance as a mixture of a temporal Hawkes process and a Neymann-Scott or log-Gaussian Cox process for the spatial part is a natural follow-up to the present work. Such a modelling could be in particular relevant if one intends to provide a short-term forecast of electrical activity.}

\section{Acknowledgements}

The authors would like to take the opportunity to thank Météorage and Météo-France, and in particular Maxime Taillardat and Olivier Mestre for providing us with the data and for scientific exchanges about these data. The research of JF Coeurjolly is funded by Labex PERSYVAL-lab ANR-11-LABX-0025.

\bibliography{references}


\newpage

\appendix
\section{Additional figures}
\label{app:figures}

\subsection{Exploratory data analysis}\label{app:eda}

\begin{figure}[htbp]
\centering
\includegraphics[width=.8\textwidth]{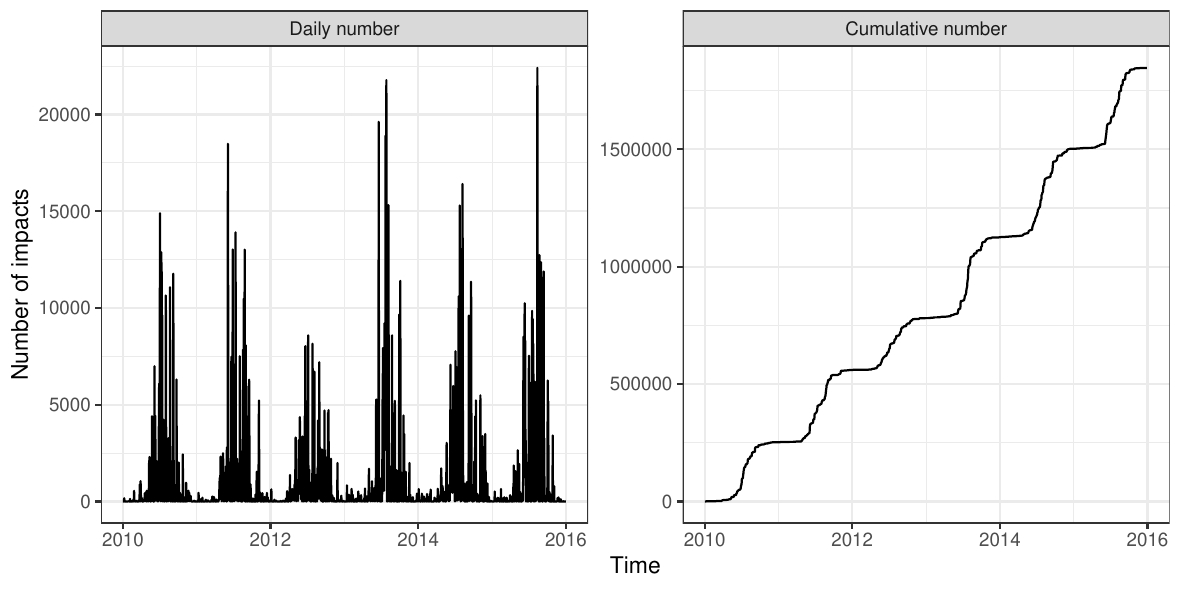} 

\bigskip

\includegraphics[width=\textwidth]{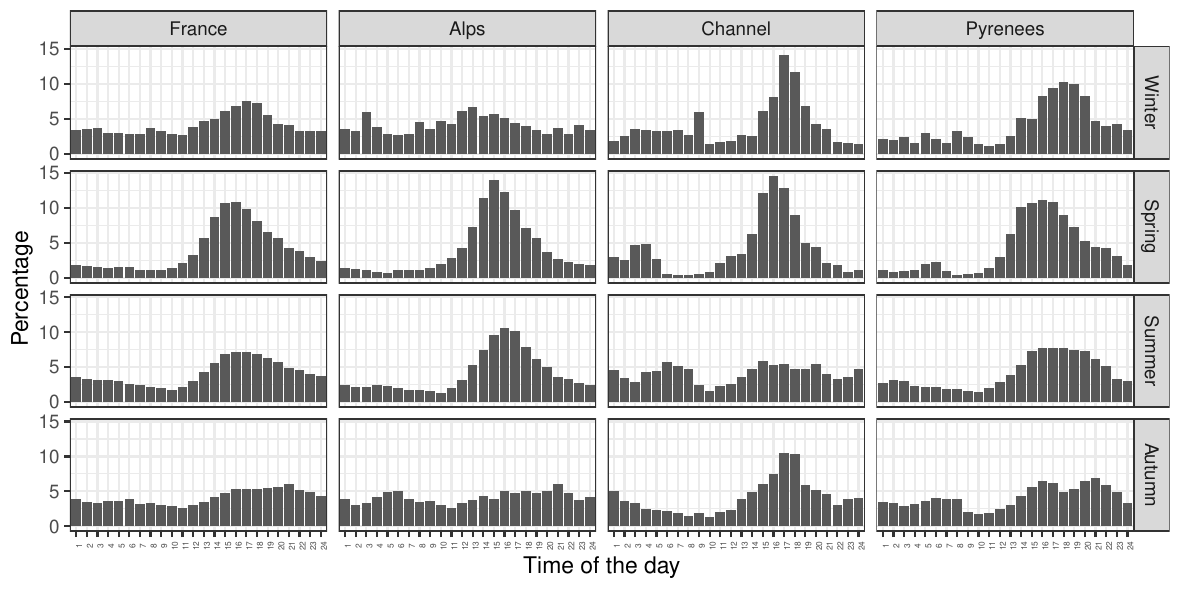}
\caption{(Top) Daily counting process (left) and cumulative counting process (right) of impacts numbers over France from 2010 to 2015. (Bottom) Distribution of impacts in terms of the time of day for each season over France and Alps (first two columns on the left) and over the Channel and Pyrénées (last two columns on the right). 
}\label{fig:countingProcess}
\end{figure}


\begin{figure}[htbp]
 \centering 
\includegraphics[width=.9\textwidth]{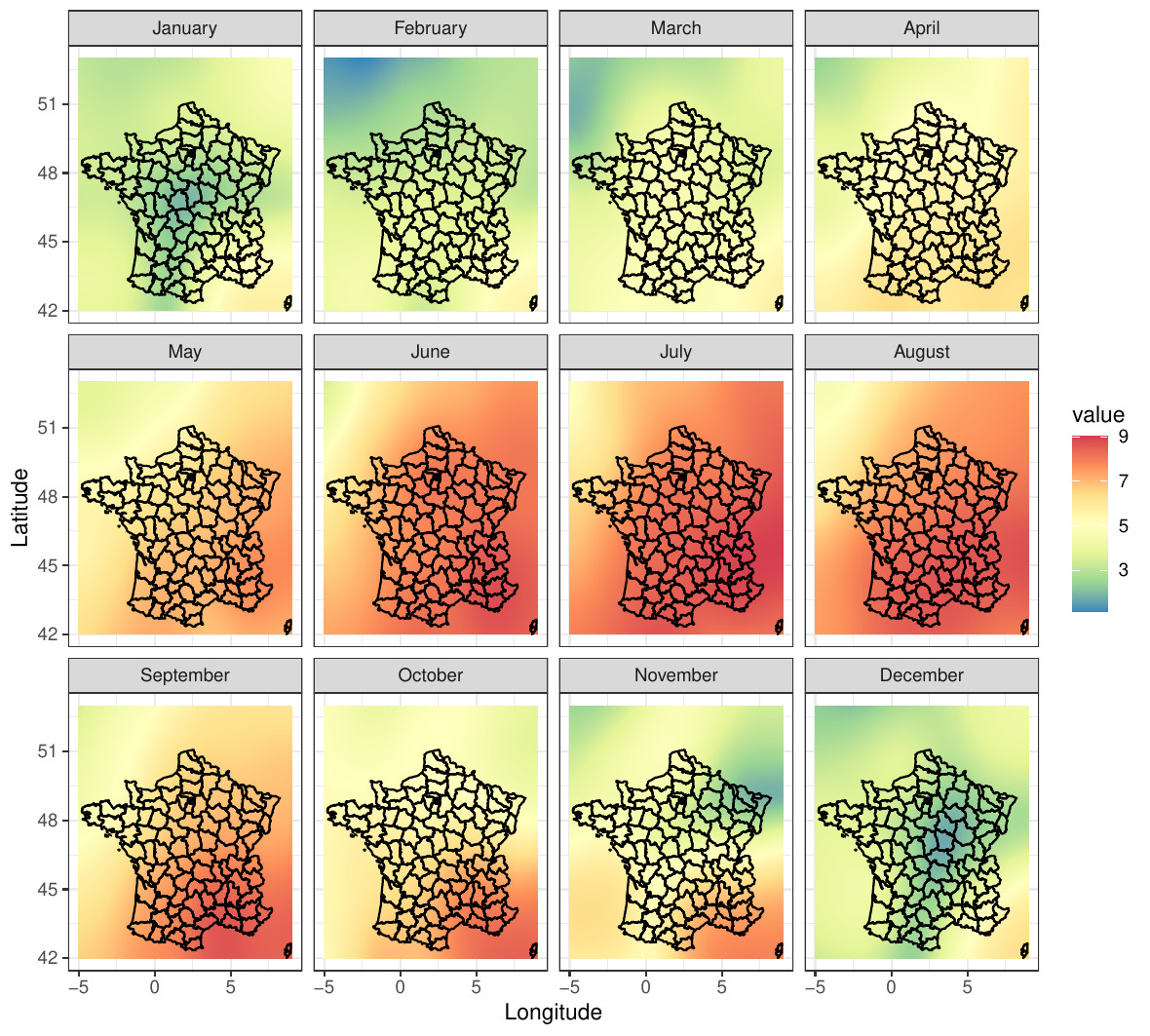}
 \caption{Logarithms of non-parametric kernel spatial intensity estimates of the intensity of impacts aggregated by month.}
  \label{fig:impacts-mois}
\end{figure}

\begin{figure}[htbp]
 \centering
\includegraphics[width=.9\textwidth]{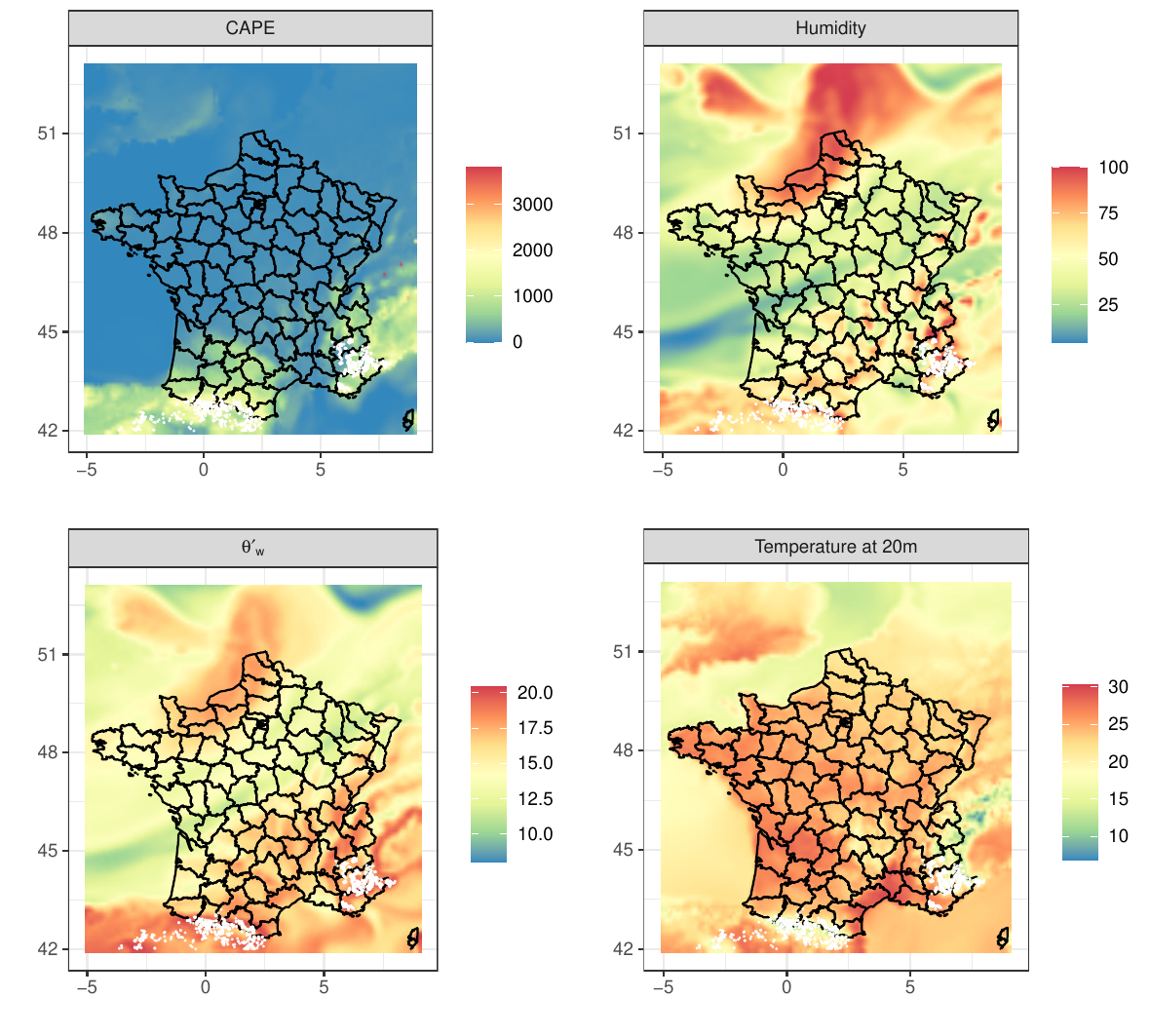}

\bigskip

\includegraphics[width=.9\textwidth]{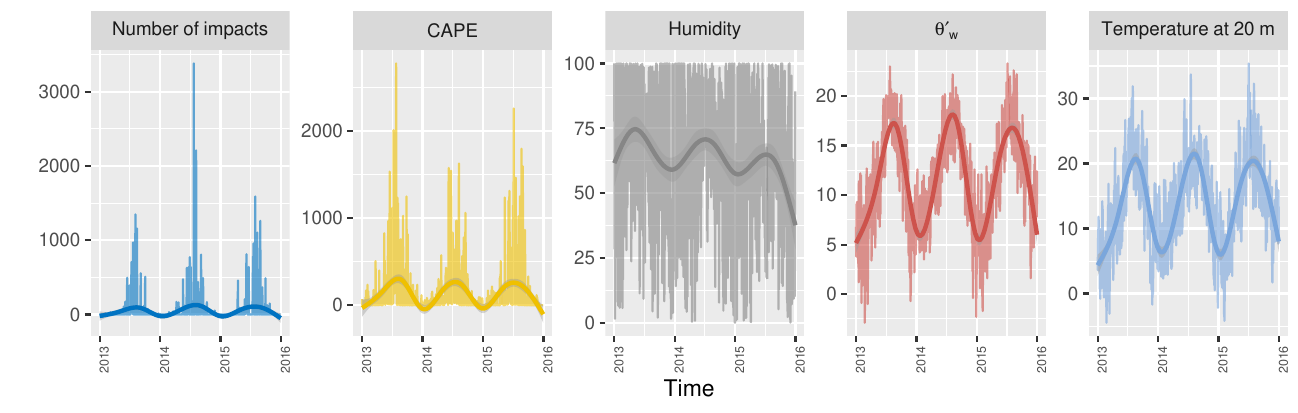}
\caption{(Top) Maps of CAPE (top left); humidity (top right); $\theta_w'$ at 850 hPa (Wet-bulb potential temperature at 850 hPa) (bottom left);  temperature at 20m (bottom right), overlaid by the locations of impacts on July 14, 2013 over the period 12:00-18:00. (Bottom, from left to right) Numbers  of impacts over the period 12:00-18:00 for each day from 2013 to 2015, around location $x_0$ (defined in Figure~\ref{fig:introductionData}); and covariates at 12am at location $x_0$,  in terms of time  (CAPE; Humidity; $\theta_w'$; Temperature at 20m).}
  \label{fig:covar}
\end{figure}

\begin{figure}[H]
 \centering 
\includegraphics[width=.9\textwidth]{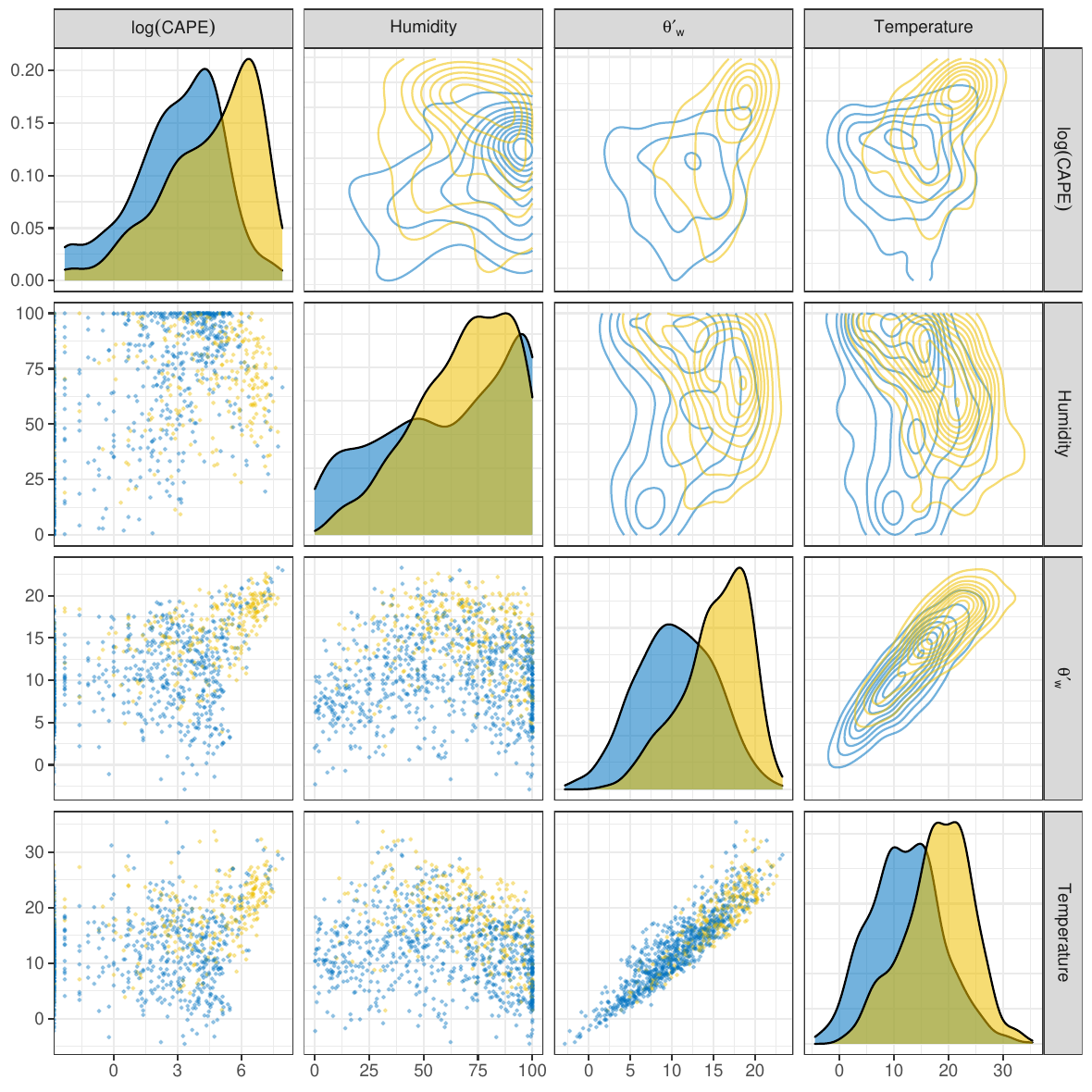}
\caption{Pairplots (lower triangular part), marginal distribution (diagonal), and bivariate distributions (upper triangular part) of log(CAPE) (0 values for CAPE are omitted), Humidity, $\theta^\prime_w$, Temperature at 20m around location $x_0$ at 12 am. Yellow (resp. blue) color corresponds to times for which at least one (resp. no) lightning strike is observed in the corresponding cell during the period 12:00-18:00. The time period spans from 2013 to 2015.}
  \label{fig:pairplots}
\end{figure}



\subsection{Simulation study}\label{app:simus}

\begin{figure}[H] 
\centering
\includegraphics[width=.9\textwidth]{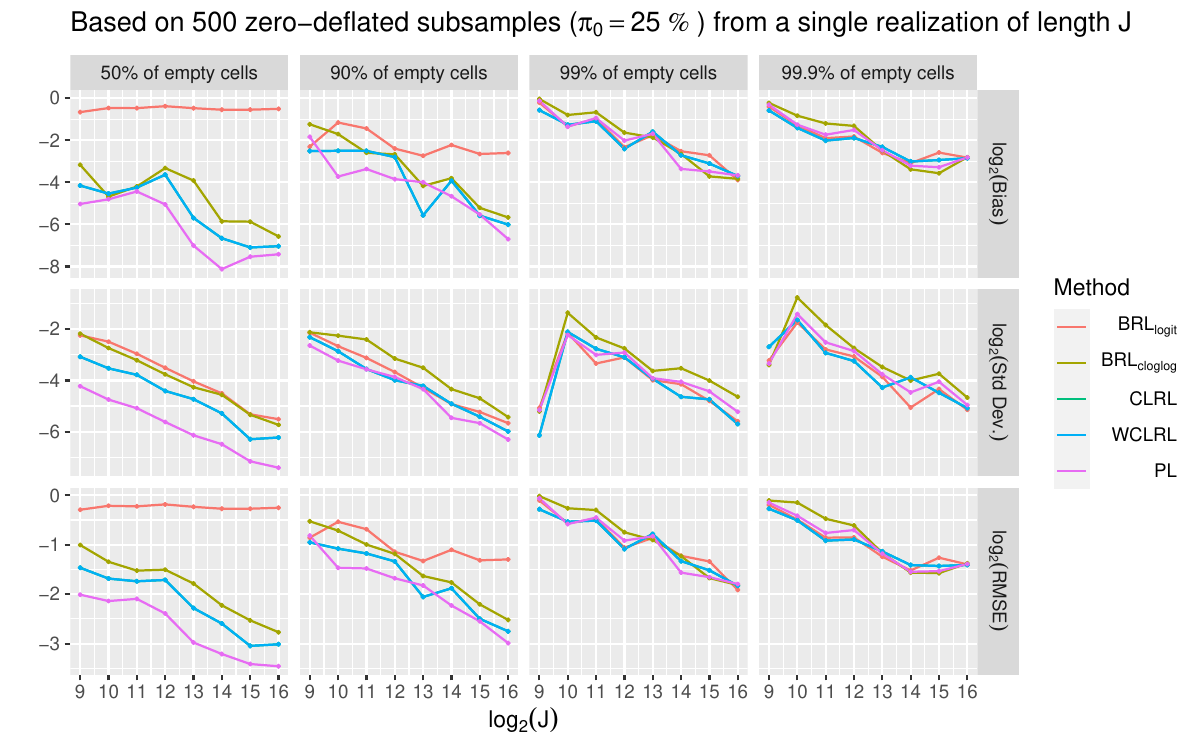}
\includegraphics[width=.9\textwidth]{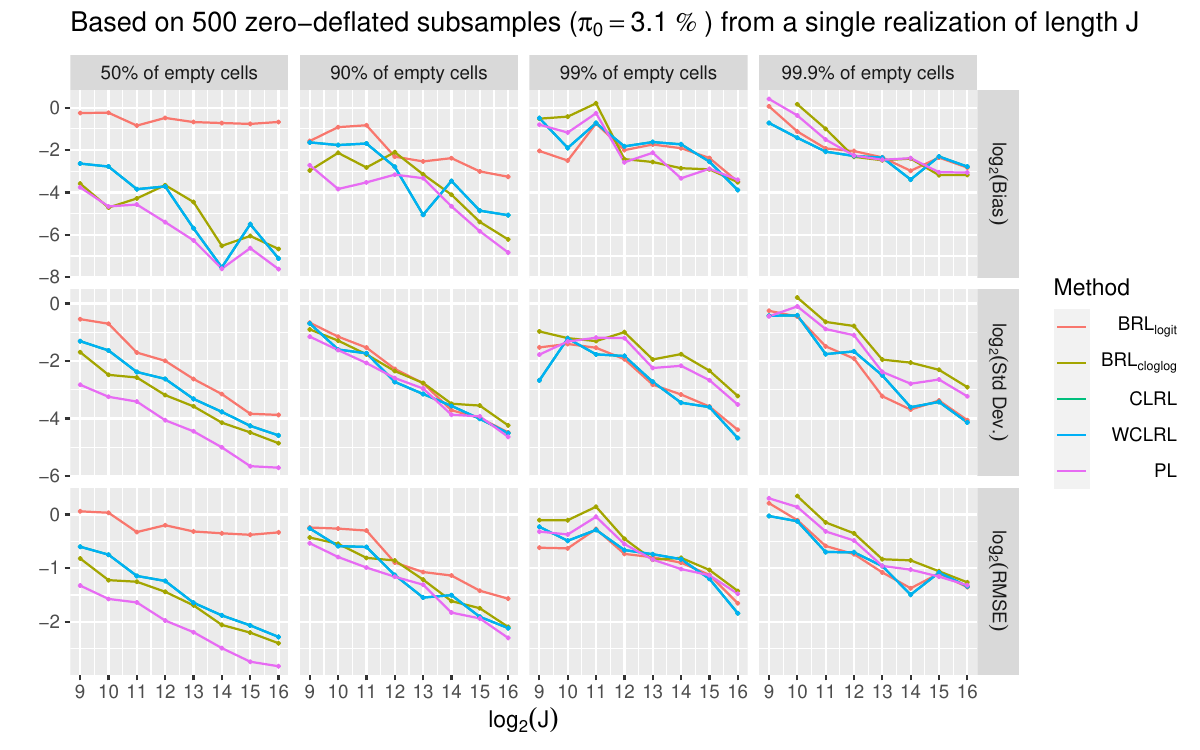}
\caption{Description similar to Figure~\ref{fig:simulation} with $\pi_0=25\%$ (top) and $\pi_0=3.1\%$ (bottom).}\label{fig:simulation2}
\end{figure}


\end{document}